\documentclass[twocolumn]{aastex63}

\usepackage{amsmath}
\usepackage{amssymb}
\usepackage{braket}
\usepackage{amssymb}
\usepackage{color}
\usepackage{xcolor}
\usepackage{graphicx}
\usepackage{latexsym}
\usepackage{listings}
\usepackage{mathrsfs}
\usepackage{letltxmacro}
\usepackage[normalem]{ulem}

\usepackage{subfigure}

\usepackage{verbatim}
\usepackage{tabularx}
\usepackage{ragged2e}
\usepackage{multirow}
\usepackage{amsbsy}

\usepackage{numprint}
\usepackage{makecell}
\usepackage[utf8]{inputenc}

\newcommand{\R}{\mathbb{R}}
\newcommand{\N}{\mathbb{N}} 

\newcommand{\V}{\mathcal{V}}

\newcommand{\F}{\mathcal{F}}

\newcommand{\M}{\mathcal{M}}

\newcommand{\I}{\mathcal{I}}

\newcommand{\B}{\text{B}}

\npdecimalsign{.}
\npthousandsep{,}

\newcolumntype{Y}{>{\RaggedRight\arraybackslash}X}

\graphicspath{{./figures/}}

\sloppypar

\shorttitle{Binary Progenitors of GRBs}
\shortauthors{Lloyd-Ronning}
\graphicspath{{./}{figures/}}

\begin{document}

\title{ Radio Loud vs. Radio Quiet Gamma-ray Bursts: the Role of Binary Progenitors.}

\correspondingauthor{}
\email{lloyd-ronning@lanl.gov}

\author[0000-0003-1707-7998]{Nicole Lloyd-Ronning}
\affiliation{CCS-2, Computational Physics and Methods, Los Alamos National Lab, Los Alamos NM 87544}
\affiliation{Department of Math, Science and Engineering, University of New Mexico, Los Alamos, Los Alamos NM 87544}



\begin{abstract}

 We explore the possibility that radio loud gamma-ray bursts (GRBs) result from the collapse of massive stars in interacting binary systems, while radio quiet GRBs are produced by the collapse of  single massive stars.  A binary collapsar system can have the necessary angular momentum and energy budget to explain the longer prompt gamma-ray durations and higher isotropic energies seen in the the radio loud sub-sample of long GRBs.  Additionally binary systems can lead to rich and extended circumstellar environments that allow for the presence of the long-lived radio afterglows seen in the radio loud systems.  Finally, the relative fraction of stars in binary systems versus single star systems appears consistent with the fraction of radio loud versus radio quiet GRBs. \\ \\

\end{abstract}




\section{Introduction}


  The connection between long gamma-ray bursts (those GRBs with prompt gamma-ray durations lasting longer than a few seconds) and the deaths of massive stars is strongly supported, both theoretically and observationally, including the direct association of  supernova events coincident with many long gamma-ray bursts \citep{Woos93,MW99, BKD02, Hjorth03,WB06,WH06,KNJ08a, KNJ08b, HB12, Ly17}. It is also clear that long gamma-ray bursts (lGRBs) are rare events - not every massive stellar death produces a highly relativistic jet that leads to a gamma-ray burst and its subsequent afterglow. The conditions required to produce an lGRB from a collapsing star \citep{MW99,YL05,HMM05,Yoon06,WH06}, and indeed the jet launching mechanism itself  \cite[e.g.][]{BK08, KB09, LR19bz,KP21}, are still open questions. It is an ongoing pursuit to determine and understand {\em which} stars make gamma-ray bursts, and {\em why}.
  
  We can hope to unravel this mystery through a careful examination of the emission we observe from these objects.  For example, the $10$'s to $100$'s seconds of highly variable prompt gamma-ray emission offers insight into the central engine of these objects \citep{KPS97, NP02, LR16,LR18}.  The duration and strength of this prompt emission, produced by internal dissipation processes in the jet, reflect the amount of angular momentum and the overall energy budget of the black hole-accretion disk system.\footnote{Here, we do not consider magnetar central engines, although our arguments can be generalized to that case.}  Meanwhile, the long-lived (days, weeks, months, years) afterglow of a gamma-ray burst, produced by the deceleration of the jet as it sweeps up the external medium, carries information about the circumburst environment since it is the cumulative mass of this environment that serves to decelerate the jet.
  
  Unfortunately, unraveling the properties of a GRB progenitor from the prompt and afterglow light curves and spectra is difficult, while the analysis is plagued with degeneracies among the model parameters;  that is, several equally reasonable models may fit the data equally well and it becomes difficult to distinguish among them \cite[e.g.][]{Li07,Via18,TLR21}.  As more and better broadband data arrive, and statistical fitting and analysis techniques progress, our understanding is improving (see, e.g., the paper by \cite{BGG20}, who discuss ways to break the degeneracies by considering the shape or curvature of the lightcurve).  However, we still have much to learn about teasing out the underlying physics of a GRB.
  
  Another approach is to search for correlations between observed variables, which can offer clues into physical processes playing a role in the GRB event.  For example, \cite{LPM00} found that the well known gamma-ray flux-spectral peak energy correlation seen in BATSE GRB data \citep{Mal95} is due to an intrinsic correlation between isotropic radiated energy $E_{iso}$ and the peak of the intrinsic spectrum $E_{p}$; they predicted a relationship $E_{iso} \sim E_{p}^{0.55 \pm 0.15}$, which was indeed later found in the data for those lGRBs with measured redshifts \citep{AM02}.   \cite{LPM00} attributed this correlation to synchrotron radiation as the dominant emission mechanism producing the prompt gamma-rays.  Many other significant correlations have been seen in the GRB data, all of which offer clues into the physics at play, and which can potentially be linked to the underlying progenitor system (for an extensive and detailed summary of relations seen in GRB data and their physical implications, see the book by \cite{Dai19} and the following reviews: \cite{Dai17, Dai18, DA18}).
  
  One can also examine whether the distributions of different observed variables suggest different classes, or different potential progenitors, of GRBs.\footnote{A classic example of this is the distribution of observed prompt gamma-ray duration \citep{Kouv93}.  This distribution is highly bimodal and it was suggested early on that long GRBs ($>2s$) come from different progenitors than short GRBs ($<2s$).  This indeed has been confirmed thanks to coincident gravitational wave emission indicating that short GRBs result from double neutron star mergers \citep{Ab17}.}  Along these lines, inspired by the presence of radio-loud and radio-quiet AGN, \cite{HGM13} first suggested that there may be a population of intrinsically radio quiet GRBs - GRBs for which very little radio emission is intrinsically present.  \cite{LR17} and \cite{LR19} tested this idea and, looking in detail at the data and accounting for observational selection effects, found distinct differences between GRBs with and without radio afterglows. Those with radio afterglows are significantly longer in their prompt duration and more energetic than those without, potentially suggesting two different populations.
 
   In this paper, we consider the possibility that radio loud lGRBs come from the collapse of a massive star that is in an interacting binary system.  The binary interaction allows for: 1) the high angular momentum needed to create a long-lived accretion disk and launch a jet,  2) a potentially higher energy budget both from the jet power's dependence on angular momentum and mass gained from a companion, and 3) a dense and extended circumburst environment, in which the GRB jet is more likely to produce the bright radio afterglow.  Additionally, we consider the fraction of stars thought to be in binary systems and compare this to the fraction of radio loud GRBs.

   An lGRB occurring from the collapse of a massive star in a binary system has been considered in the past \cite[e.g][]{Iv02,vdHy07,BK10,Pod10, CSE20}.  In particular, \cite{vdHy07} and \cite{BK10} consider a massive star collapse in a close binary system with a compact object companion.  The latter show that the GRB progenitor can have significantly higher mass compared to a single star system, sufficient spin to launch a GRB jet, and in some cases a very long lived accretion disk. \cite{CSE20} consider binary progenitors for long gamma-ray bursts and show that not only can binary systems provide the necessary stripping of the stellar envelope (see also \cite{Pod10, Lap20, FW22}) and sufficient angular momentum to launch a jet, but also, from their population synthesis models, can broadly accommodate the rates of lGRBs. 
   
   Recently, \cite{Bav21} examined massive stars collapsing in binary systems where both stars end their lives as black holes (and therefore are the progenitors of binary black hole mergers observed by LIGO and Virgo). Modelling three types of binaries - a) close binaries with a common envelope phase, b) wider binaries with stable mass transfer, and c) close binaries with chemically homogeneous evolution - they found that a star collapsing in systems corresponding to cases a) and c) can produce a highly spinning black hole that launches a GRB jet. Furthermore, their rate estimates, based on LIGO/Virgo's binary black hole gravitational wave observations, are consistent with these systems producing a large fraction (the majority) of lGRBs.

\begin{figure*}[!t]
	\includegraphics[width=2.3in]{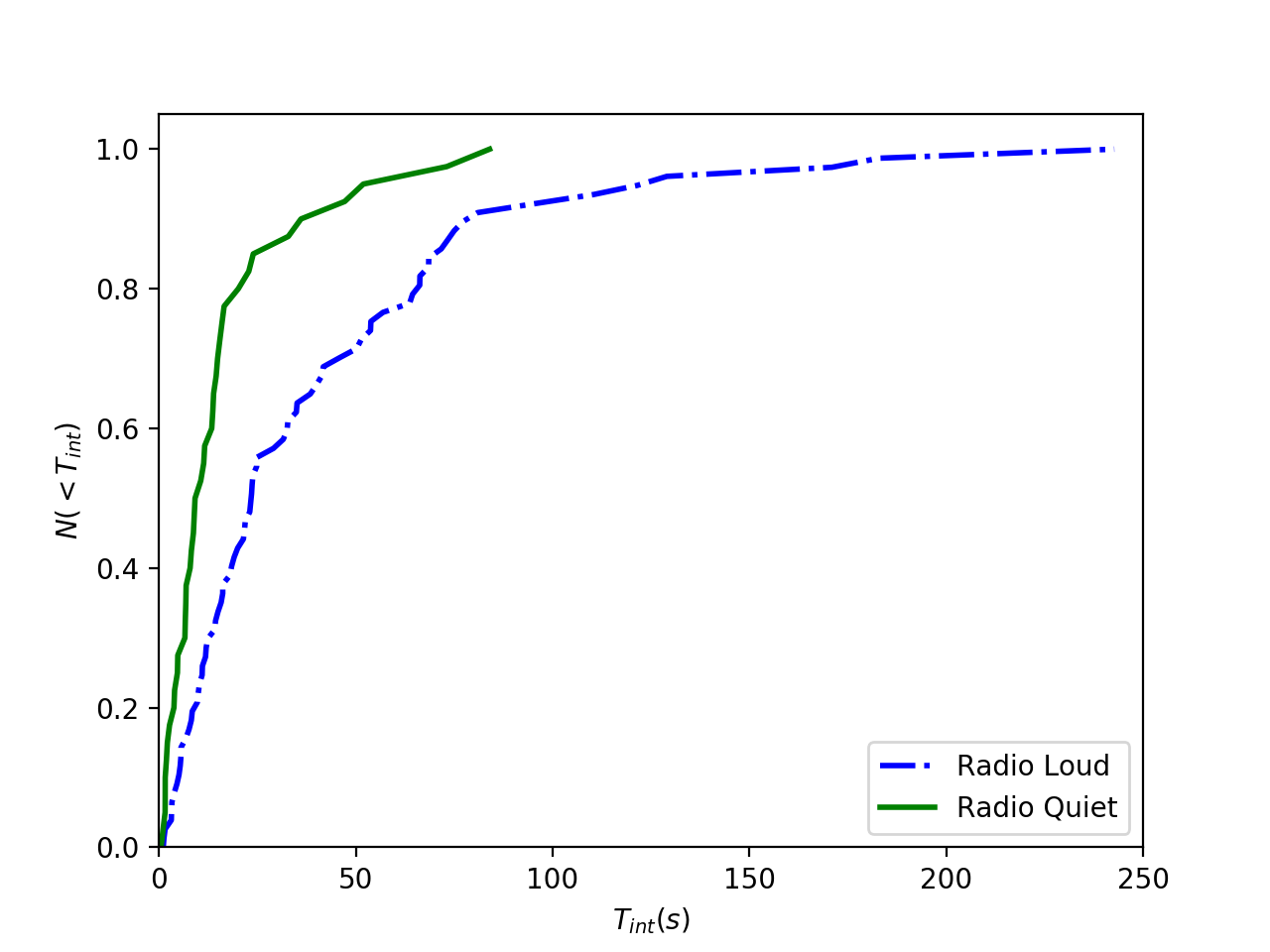}\includegraphics[width=2.3in]{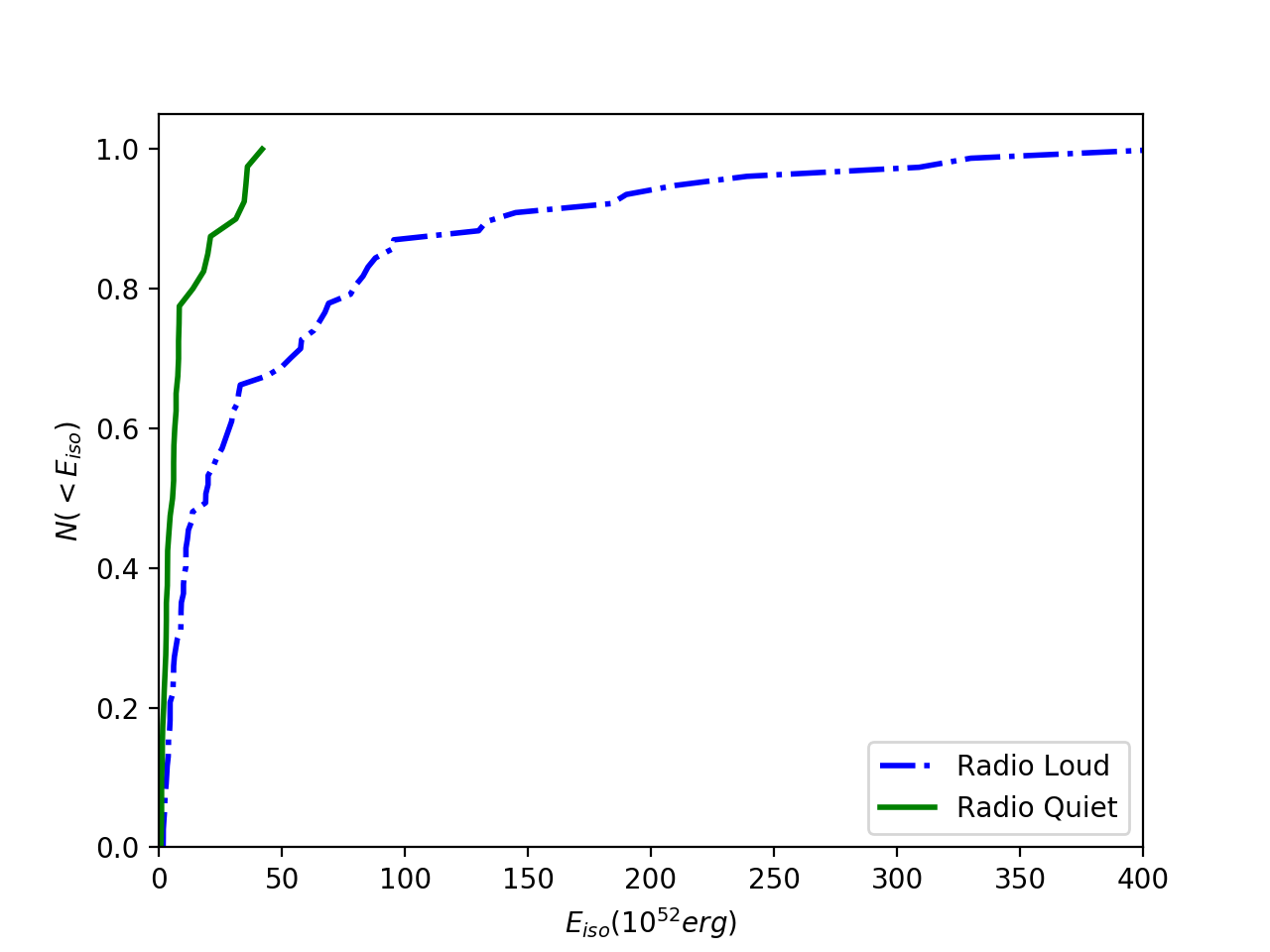}\includegraphics[width=2.3in]{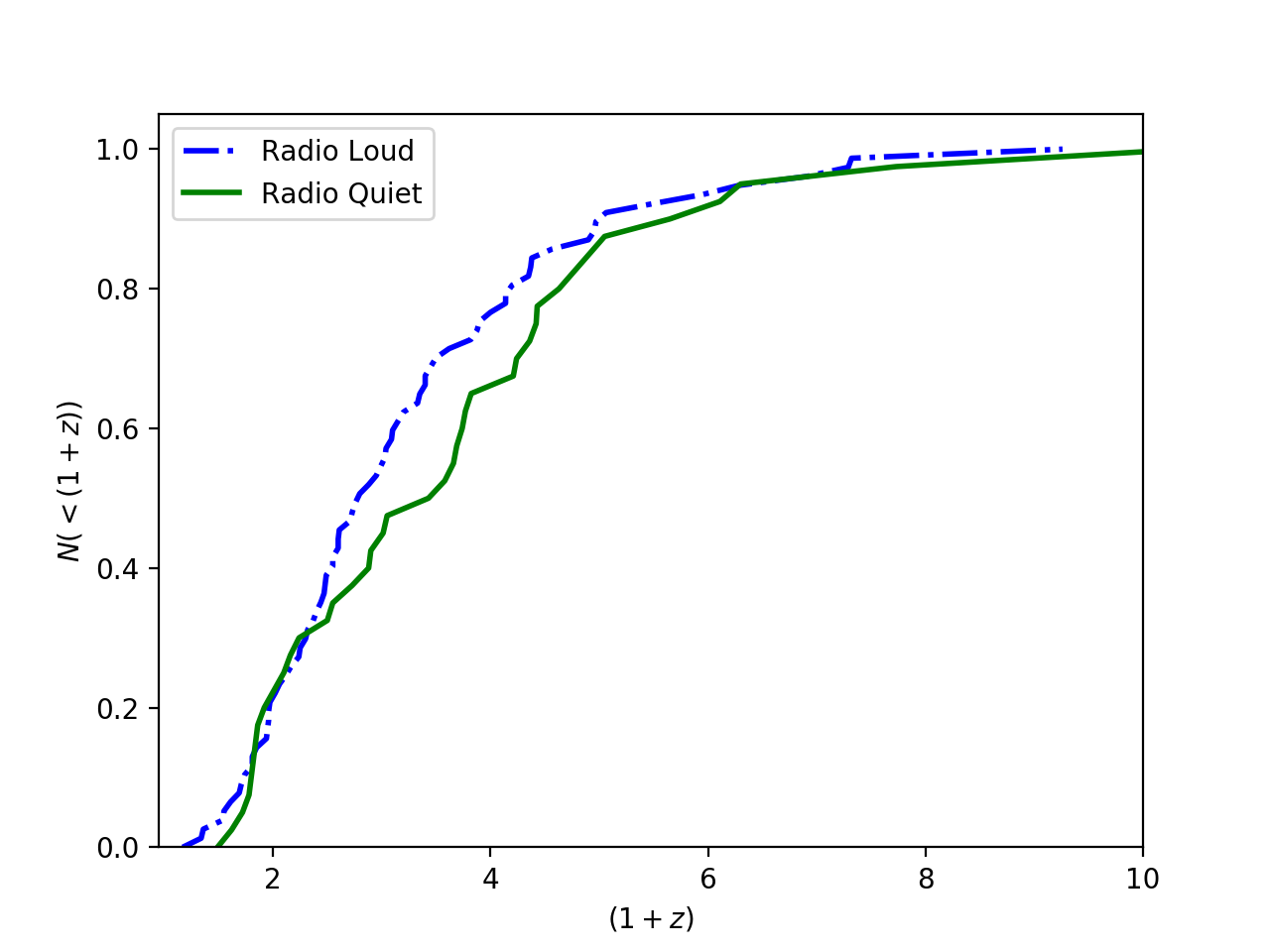}\\
	\includegraphics[width=2.3in]{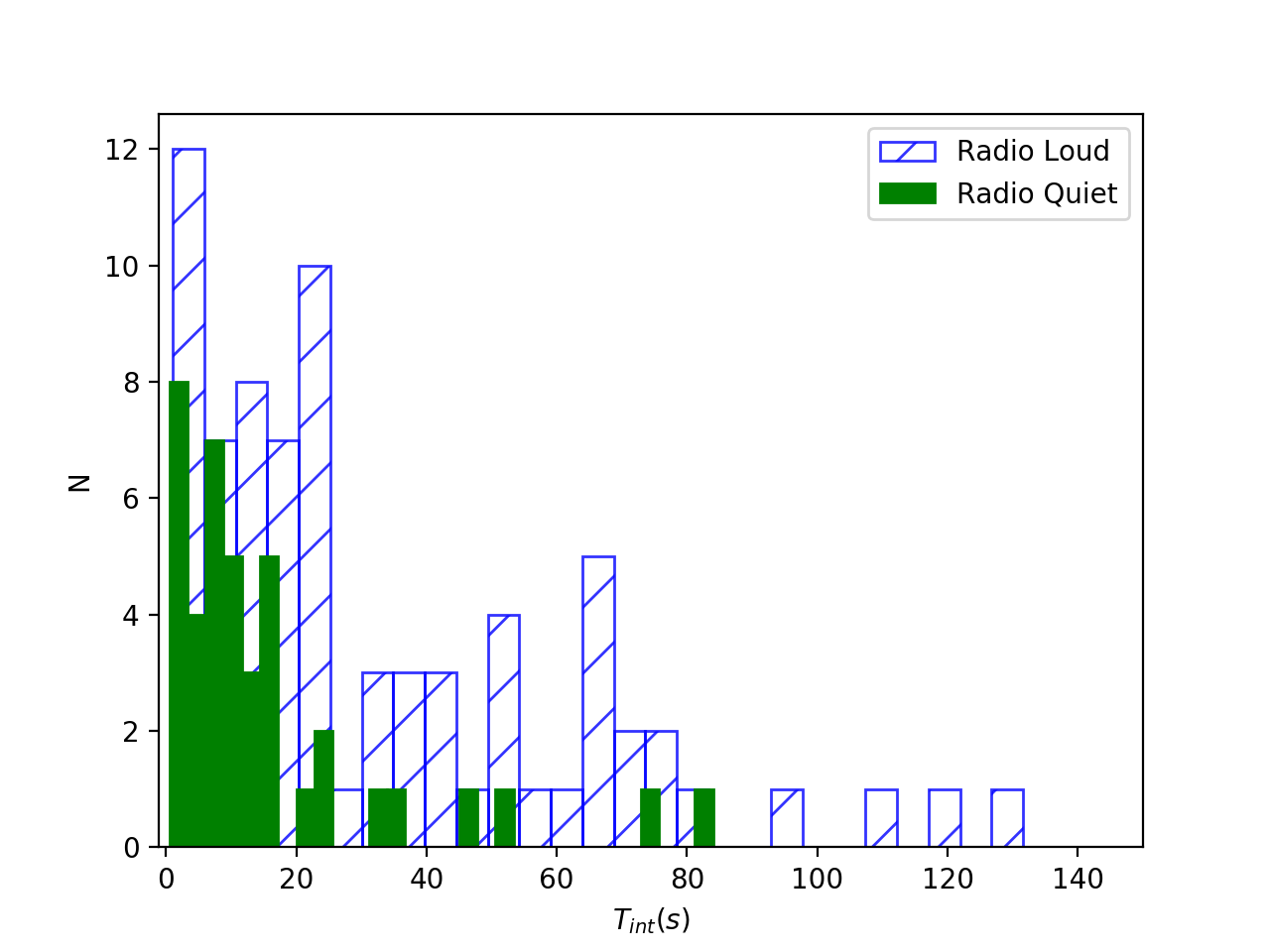}\includegraphics[width=2.3in]{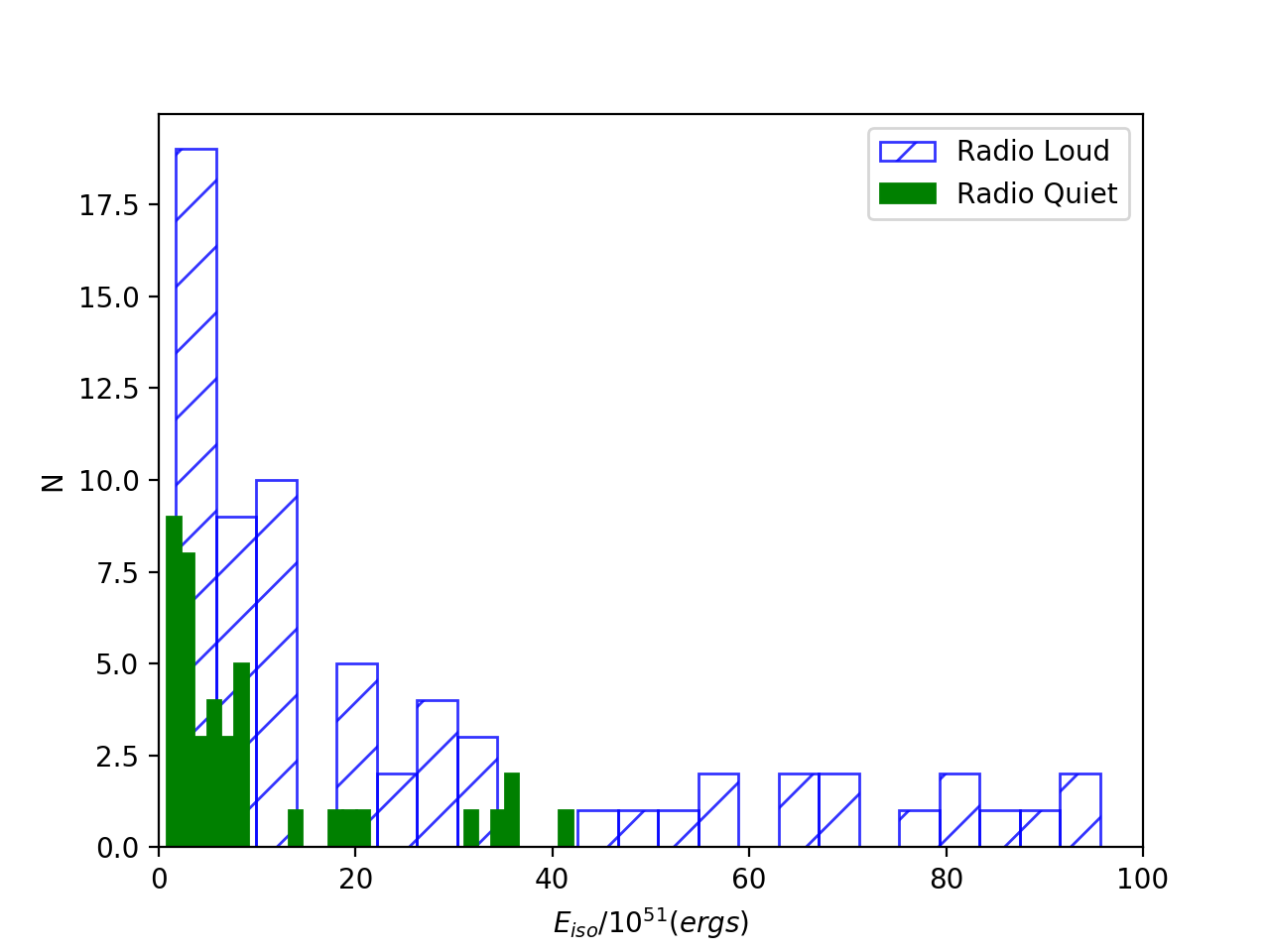}
    \includegraphics[width=2.3in]{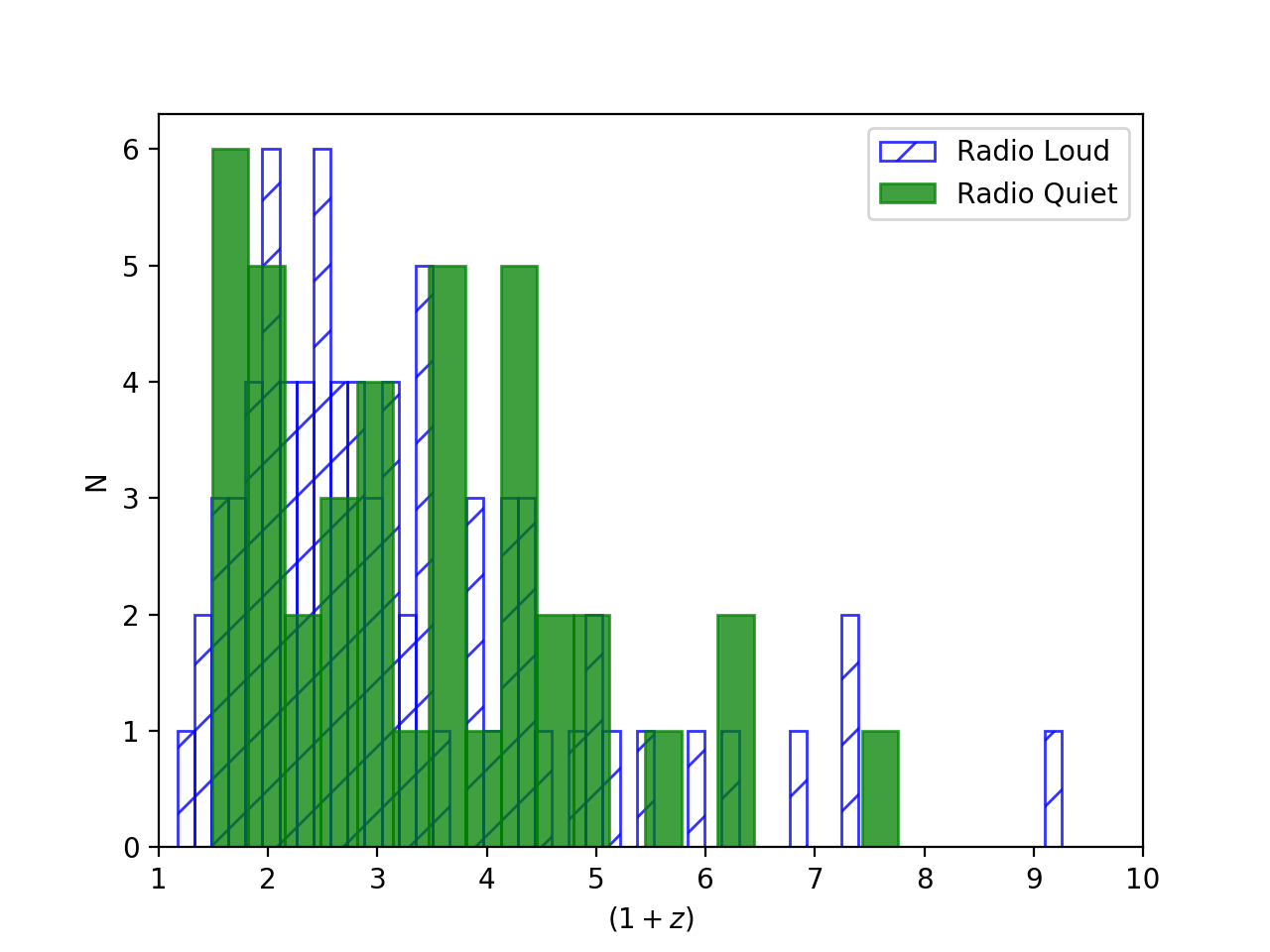}
    \caption{Cumulative (upper row) and differential (lower row) distributions of intrinsic GRB duration $T_{int}$ (left panel), isotropic equivalent energy $E_{iso}$ (middle panel), and redshift $(1+z)$ (right panel) for radio bright (dotted blue line) and radio quiet (solid green line) GRBs.  A KS test indicates the radio loud and quiet samples have significantly different distributions in both $T_{int}$ and $E_{iso}$, but not $(1+z)$. From \cite{LR19}}
    \label{fig:cumdist}
\end{figure*}

In this paper, we argue that radio loud lGRBs are a result of the collapse of a massive star in an interacting binary system, while the radio quiet GRBs have single star progenitors (or exist in non-interacting binaries or multiples).  Our paper is organized as follows: In \S 2, we summarize the observational results that suggest that radio loud and radio quiet lGRBs come from  distinct populations. In \S 3, we discuss how binary interaction can affect the duration, energy budget and circumburst medium of an lGRB, and accommodate the trends we see in the radio loud and radio quiet data.  In \S 4, we discuss the rates of binary systems and show that lGRBs with radio afterglows are consistent with these rates.   In \S 5, we present our conclusions.

\begin{deluxetable}{lccc}
\tablecaption{Radio Bright vs Radio Dark Samples}
\tablecolumns{4}
\tablewidth{\linewidth}
\tablehead{Sample & $\bar{z}$ & $\bar{T}_{int}/s$ & $\bar{E}_{iso}/10^{52}$ erg}
\startdata
 Radio Bright (78 bursts) | & 2.8 & 39. & 51. \\
 Radio Dark (41 bursts) | & 2.6 & 16. & 9. \\
\enddata
\tablecomments{ Average values of the redshift $z$, intrinsic prompt duration $T_{int}$, and isotropic emitted energy $E_{iso}$ for the sample of energetic ($E_{iso} > 10^{52}$ erg) bursts with and without detected radio afterglows. A Student's $t-$test gives a probability of less than $.001$ that these samples are from the same distribution, based on the average values of $T_{int}$ and $E_{iso}$.}
\label{}
\end{deluxetable}

\section{Properties of Radio Loud versus Radio Quiet GRBs}
Based on the suggestion by \cite{HGM13} that there may exist a truly radio dark population of GRBs, \cite{LR17} and \cite{LR19} analyzed a sample of energetic lGRBs for which radio follow-up was attempted.  Their data are primarily taken from \cite{CF12}, although their second paper adds to the sample utilizing data found on Jochen Greiner's GRB page\footnote{\url{2 http://www.mpe.mpg.de/ jcg/grbgen.html}}.  After taking steps to account for detector selection effects, they found somewhat suprising results: radio bright GRBs tend to be significantly longer (by more than a factor of 2) in their prompt gamma-ray duration and have significantly higher isotropic energy than radio dark GRBs (we note they found no correlation between isotropic energy and measured radio flux in the radio bright subsample, suggesting that this is not simply a flux-limit effect). Importantly, the redshift distributions of the radio bright and radio dark samples are {\em not} statistically different. These results were recently confirmed by \cite{Zhang21}, and summarized in Table 1.

 
 \begin{figure}
    \centering
    \includegraphics[width=0.5\textwidth]{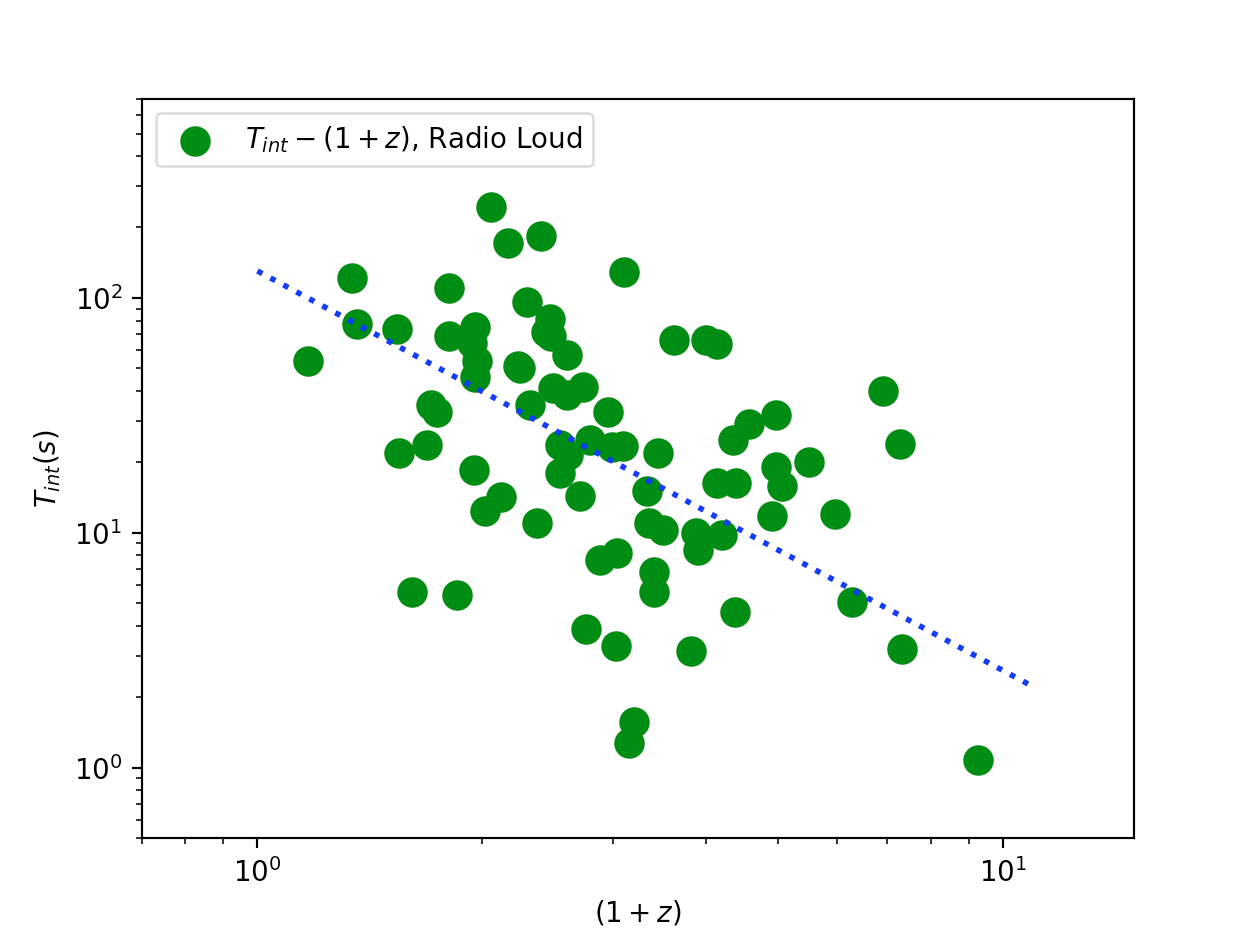}
    \caption{Intrinsic duration versus redshift for the radio bright sample. Even when accounting for selection effects that artificially truncate the upper right and lower left of this plane, we find a statistically significant ($> 4 \sigma$) anti-correlation.  The correlation is not present in the radio quiet sample.}
    \label{fig:my_label}
\end{figure}

 \cite{LR19} also found that the prompt duration for the radio bright sample is anti-correlated with redshift in only the radio bright sample (see Figure 2; this was confirmed by \cite{Dain21} in a sub-sample of radio bright GRBs with plateau-like features). Additionally, they found that {\em extended} very high energy emission (VHE; above 100MeV) is present only in the radio bright sample. These results require further confirmation, but are also suggestive of two different sources/environments for the radio bright and dark sub-samples. For example, \cite{Huang21} showed that strong wind environment suppresses VHE emission (and/or it decays faster) and the presence of VHE is more likely in a ISM-like medium.
 We note they found no apparent difference between X-ray and optical emission between the radio bright and dark samples, nor in their jet opening angles; a more detailed analysis of the differences between radio bright and dark samples is in progress (Chakraborty et al., in prep).  
 
  Therefore, because observables related to both the inner engine (duration and isotropic energy) and environment (radio flux) of the GRB appear to be significantly different between the radio quiet and radio loud samples, these results suggest that these populations could be coming from different progenitor systems. Here, we consider in broad terms whether the radio loud sample may be a result of a GRB occurring in an interacting binary system.\footnote{We emphasize that we are considering the collapse of a star in a binary system in which interaction with the companion is significant, but not necessarily a merger as the catalyst for the GRB.} \\ \\
  
\section{General Characteristics of GRBs in Binary vs Single Star Systems}  

 We consider the expected prompt duration, energy budget and circumburst medium profile of a massive star in an interacting binary system.  \\
 
\subsection{Angular Momentum and GRB Duration}
  The details of the particle acceleration and radiation processes that lead to the highly variable prompt gamma-ray emission of a GRB are still a matter of investigation, but it is believed this emission results from {\em internal} dissipation processes, like internal shocks or magnetic reconnection, in the jet as opposed to an external shock \cite[for a comprehensive review summarizing the arguments for this, see][]{Pir04}.  That means the duration of the gamma-ray emission is constrained by the lifetime of the jet, which itself depends on the amount of time the accretion disk sufficiently feeds the black hole and/or the jet launching mechanism is active.  One can estimate this timescale by considering the amount of material supplied to the central engine (i.e. the amount of mass in the disk) divided by the average rate at which that mass is accreted on to the central black hole. 
  
  \begin{equation}
  T_{int} \approx M_{{\rm disk}}/\dot{M}_{\rm disk}
  \end{equation}
\noindent where $T_{int}$ is the intrinsic prompt gamma-ray duration, $M_{{\rm disk}}$ is the mass in the disk and $\dot{M}_{\rm disk}$ is the average accretion rate.  

   These variables - the amount of mass in the disk and the (complicated, ever-changing) accretion rate are difficult to determine.  A disk will form only if there is enough specific angular momentum, $l$, in the gas (angular momentum per unit mass), $l > l_{critical} \sim 2R_{g}c$, where $R_{g}$ is the gravitational radius of the central black hole \cite[see, e.g., the descriptions in][]{PB03,Pr06,JP08}, and a system with more angular momentum has the ability to create a more massive disk\footnote{Again, here we consider a a black hole central engine but see, e.g., \cite{KB07} and \cite{Zou19} for discussion of a magnetar central engine.}.  That is, more of the collapsing stellar mass can be centrifugally supported in a disk as opposed to collapsing directly to the central black hole\footnote{ \cite{JP08} showed this mass can be estimated as $M_{disk} \approx l_{max}c/4G - M_{core}$, where $l_{max}$ is the maximum angular momentum in the collapsing gas and $M_{core}$ is the core mass of the collapsing massive star.}.   
   
   Additionally, the accretion rate is in general lower for higher angular momentum systems - a more highly spinning disk will accrete at a lower rate than a lower angular momentum disk \cite[for a look at this effect in both stellar mass and supermassive black hole-disk systems, see, e.g.,][ respectively]{PB03, RG15}.  As is clear from equation 1 above, both of these effects of a higher angular momentum system - that is, a more massive $M_{disk}$ and lower accretion rate $\dot{M}_{disk}$ - serve to increase the duration, $T_{int}$, of lGRB.\\ 
   
   We can make an order of magnitude estimate of how duration depends on the amount of angular momentum in our GRB system as follows:  We assume the GRB central engine is a black hole-accretion disk system, which serves to launch the GRB jet. 
   For a solid rotating disk, we can write the mass of the disk as:
   \begin{equation}
       M_{\rm disk} \sim 2 J R_{disk}^{-2} \omega^{-1}
   \end{equation}
   
   \noindent where $J$ is the angular momentum, $R_{disk}$ is the radius of the disk and $\omega$ is its average rotational velocity.  A rough but informative estimate of the accretion rate of the disk is given by the mass in the disk divided by a dynamical timescale, $\dot{M} \sim M_{\rm disk}/t_{\rm dyn}$.  For a disk with angular momentum $J$ and mass $M_{\rm disk}$ rotating around a black hole of mass $M_{BH}$ at an average rotational velocity of $\omega$, the dynamical time can be written as:
   \begin{equation}
       t_{dyn} \sim J^{3/4}M_{\rm disk}^{-3/4}\omega^{-3/4}G^{-1/2}M_{BH}^{-1/2}
   \end{equation}
   \noindent where $G$ is Newton's gravitational constant.  Then the accretion rate can be estimated as:
   \begin{equation}
       \dot{M} \sim J^{-3/4}\omega^{3/4}M_{\rm disk}^{7/4}M_{BH}^{1/2}G^{1/2}
   \end{equation}
   \noindent so that the intrinsic duration of the gamma-ray burst, which - again - is given by the lifetime of the disk that sustains the GRB jet, is related to angular momentum as:
   \begin{equation}
       T_{int} \approx M_{{\rm disk}}/\dot{M}_{\rm disk} \propto J^{7/4}
   \end{equation}
   
   This simplistic order-of-magnitude estimate aligns with the basic intuition that a system with more angular momentum should indeed lead to a longer duration GRB.  We now discuss angular momentum in single star vs. binary systems. \\
   
   \subsubsection{Prompt Duration in Single Star Systems}
   Because the angular momentum of single massive stars at the end of their lives is a huge unknown \citep{Heg05,Heg06,Ros12,Sund13}, there are a wide range of possibilities for what the expected value of the intrinsic duration might be in a single massive star model. \\
   
   \cite{JP08} looked at a number of models for the distribution of angular momentum, as well as three distinct accretion scenarios (depending on how gas in the polar region is treated), for the gas in a single massive star collapse.  They show, in these single star systems, that the total mass that forms a disk is usually only a small fraction of the stellar envelope.  From the amount of material that forms a disk, and using two different accretion rate assumptions (a constant between $0.1$ - $1.0 M_{\odot}$, and a variable accretion rate based on the free fall velocity of the gas in the torus), they estimate the duration of a single star collapsar producing a GRB, and show that most collapsars can produce GRBs with durations {\em at most} of about $50s$.  Even this duration  required somewhat extreme conditions, indicating single star collapsars producing GRBs are not only indeed very unique, but can only produce lGRBs {\em on the lower end of the prompt duration distribution}, consistent with the average prompt durations ($\sim 16s$) that \cite{LR19} found for radio quiet GRBs. \\
    
\subsubsection{Prompt Duration in Binary Systems}
    In a binary system, on the other hand, we 
   may have significant angular momentum transfer from the companion star that can in principle spin up the GRB collapsar and create a higher angular momentum system.  Stars in tidally locked binaries systems will be spun up (or down) until the spin angular momentum is roughly the same as the orbital velocity \citep{Izz04}. For the spin values needed for a successful GRB (black hole spin parameters $a \gtrsim 0.5$), this requires relatively short orbital periods ($<$ 10hr), which in turn implies a compact object companion \citep{Pod04, Jan13}. Such systems have been shown to lead to GRBs that can last hundreds to even thousands of seconds \citep{BK10}.  \cite{Bav21} recently examined binary systems that are progenitors of lGRBs, and showed that two of the three formation channels they consider (common envelope and chemically homogeneous evolution) lead to systems with significant tidal interaction and angular momentum increase.  The high angular momenta of the cores of the progenitor stars in their systems are preserved until the collapse of the second massive star, even when angular momentum transport is efficient. \\  
     
    As pointed out above and discussed further below, we consider interacting binary systems in cases where the companion star is a compact object in close orbit; this companion imparts significant angular momentum to the primary star (the massive star that will collapse and form a GRB). For our analysis below, we consider the case where the compact object companion is a black hole; however, our analysis is readily extended to a neutron star or white dwarf companion as well.\\
    
    The total angular momentum in massive star-black hole binary system is given by:
    \begin{equation}
        J =  \left(\frac{Gr(M_{BH}M_{star})^{2}}{M_{tot}} \right)^{1/2} + I_{BH}\omega_{BH} + I_{star}\omega_{star}
    \end{equation}
    \noindent where $M_{tot} = (M_{BH} + M_{star})$ is the total mass of the binary system, $r$ is the semi-major axis, $I_{BH}$ and $I_{star}$ are the moments of inertia of the black hole and star respectively, and $\omega_{BH}$ and $\omega_{star}$ are the rotational velocities of the black hole and star respectively.\\

   We consider close, interacting binary systems in which tidal locking will occur on short timescales compared to the lifetime of the massive star that collapses to create the GRB.  The tidal locking timescale is given by \citep{GLAD96}:
    \begin{equation}
        t_{lock} \approx \frac{16 r^{6} \rho_{star} \omega_{o} Q}{45 G M_{BH}^{2} k_{2}} 
    \end{equation}
    \noindent where $\rho_{star}$ is the average density of the star, $\omega_{o}$ is its initial rotational velocity, $Q$ is the so-called dissipation function of the star, and $k_{2}$ is the tidal love number. Quantities like $Q$ and $k_{2}$ are highly unknown \citep{GLAD96} and usually taken to be around $\sim 100$ and of order unity, respectively.  This expression, however, is most sensitive to the orbital separation $r$ of our binary system and it becomes clear why close binary systems are necessary for tidal locking to occur on a timescale less than the lifetime of the massive star.  Taking high mass X-ray binary systems as a guiding point for the values in the expression above, we find:
    
  \begin{equation}
  \begin{aligned}
        t_{lock} \approx & 10^{3} \ {\rm yr} \ (r/10^{12}cm)^{6} (\rho_{star}/10^{3}g cm^{-3}) \\
        & (\omega_{o}/10^{-5}s^{-1}) (Q/100)  (M_{BH}/10M_{\odot})^{2} (k_{2}/0.1) 
\end{aligned}
\end{equation} \\

\noindent indicating that for closely separated binaries, tidal locking can occur on short timescales compared to the life of the massive star.

  In this case, when the system becomes tidally locked, the angular momentum of the massive star in the binary system that will create the GRB upon collapse is given by:
  \begin{equation}
      J_{star, binary} = \alpha M_{star} R_{star}^{2} \left(\frac{G M_{tot}}{r^{3}} \right)^{1/2}
  \end{equation} 

\noindent where we have estimated the moment of inertia of the star to be $I_{star} \sim \alpha M_{star} R_{star}^{2}$. Ultimately, what we are doing here is comparing the rotation rate, $\omega_{star} = \sqrt{G M_{tot}/r^{3}}$, of a massive star in a tidally locked close binary system to that of a single massive star.  Once again, the rotation rates of single massive stars can take on a range of values and are very unknown \citep{Heg05,Heg06}, but several works \citep{Ros12,Sund13} suggest periods on the order of days to tens or even hundreds of days. \\

Using the fiducial values for our binary system as above, and conservatively taking the rotation rate of a single massive star to be $\sim days$, we find
\begin{equation}
\begin{aligned}
   & J_{binary}/J_{single}  =  \omega_{binary}/\omega_{single} \\ 
   & \approx 5 \ (M_{tot}/20M_{\odot})^{1/2} (r/10^{12}cm)^{-3/2}
   \end{aligned}
\end{equation}

Because a binary system has in principle much more angular momentum available and will, in cases of close binaries with a compact companions, spin up the massive star that will make a GRB on a timescale shorter than the lifetime of the massive star, it can sustain a disk (and therefore a jet) much longer compared to a single star system. This allows for a longer duration gamma-ray burst. \\  

   To summarize, the amount of mass present in an accretion disk of a collapsing star and rate of accretion onto the black hole both fundamentally depend on the angular momentum in the system.  Although the details are complicated (and depend on the specifics of stellar structure and gas microphysical processes), in general we expect stars with larger angular momentum to be able to support more massive disks a lower accretion rate. Because massive stars in interacting binary systems can have significantly higher angular momentum than in single star systems, they can form a more massive, longer-lived accretion disk upon collapse, leading to a longer duration lGRB.\\
  

\subsection{Jet Power and Energy Budget}
 A star in an interacting binary system can gain mass from its companion throughout its lifetime.  Indeed, there is evidence that Type II SNe in binary systems end up with larger core mass than their single-star cousins \citep{BK10,Zap21}.
The right panel of Figure 1 in \cite{Zap21} shows that that core mass of a star in a binary system that produces a TypeII SNe is roughly one and half to two times larger than that of a single star.  
 
 
  Gain in mass from a companion provides an overall larger (gravitational potential energy) budget in these systems, which may explain the higher values of inferred higher isotropic energies in radio loud lGRBs.  However, these apparently larger isotropic energies may also result from the larger angular momentum in a binary system (as discussed in the previous section), leading to a more powerful jet.  Let us consider the jet power's dependence on angular momentum.  If we assume that the GRB jet is magnetically driven (launched by, for example, the Blandford-Znajek (BZ) mechanism \citep{BZ77}, which has been shown to be a more efficient launch mechanism for an lGRB, compared to, say, neutrino annihilation), the luminosity of the jet is given by  $L_{BZ} = (kf/4\pi c) \phi^{2} \omega^{2}$, where $k$ is a geometrical factor related to the magnetic field geometry (of order $\sim 0.05$), $\omega$ is the angular velocity of the spinning black hole, $f$ is a factor of order unity, and $\phi$ is the magnetic flux on the black hole.  We can re-write this luminosity in terms of the the black hole dimensionless spin parameter $a = Jc/GM_{BH}^{2}$ ($J$ being the black hole angular momentum) and black hole mass $M_{BH}$:
 
\begin{equation}
\centering
\begin{aligned}
L_{BZ} & =  (kfc^{5}/64\pi G^{2})a^{2}\phi^{2}M_{BH}^{-2}  
\end{aligned}
\end{equation}

 The jet power depends strongly on the spin of the black hole, the magnetic flux, and the mass of the black hole, although we note the complicated interplay between all three of these variables (for example, as discussed in the previous section, the angular momentum of the system will affect how much mass initially collapses into the black hole vs forming a centrifugally supported disk).  We can simplistically write down, under the assumption of flux conservation, the magnetic flux in terms of magnetic field strength $B$ and mass of the black hole\footnote{ $\phi \approx 4 \pi B R^{2}$, where $B$ is the magnetic field, and $R$ is the Kerr radius given by $R = GM/c^{2} + \sqrt{(GM/c^{2})^{2} - a^{2}}$}, which leads us to the following expression for the observed jet luminosity \citep{mck05, TNM10, TG15, LR19bz}:
 \begin{equation}
\centering
\begin{aligned}
& L_{GRB} \approx  \\ 
 & 10^{50} {\rm erg} (\eta/0.1) (a/0.9)^{2}(B/10^{16}{\rm G})^{2}(M_{BH}/5M_{\odot})^{2}
\end{aligned}
\end{equation} 
  
\noindent and where we have used an efficiency factor $\eta$ between the BZ jet power and the observed jet luminosity, $L_{GRB} = \eta L_{BZ}$. The measured isotropic energy is a measure of the integral of this luminosity over the duration of the gamma-ray emission (the frequency at which an lGRB in general emits most of its energy).  Therefore, more highly spinning, massive systems can lead to higher measured isotropic energies, as seen in the radio loud lGRB sample\footnote{The relationship between a massive star's angular momentum and the final resultant spin of the black hole is certainly not straightforward \citep{Heg06}. We are operating under the reasonable assumption, however, that stars with higher angular momentum will on average lead to black holes with higher spin values.}.\\ 

Because the true energy emitted in the jet, $E_{\gamma}$, is related to the isotropic energy as $E_{\gamma} = \frac{1}{2}(1-{\rm{cos}}(\theta_{j}))E_{iso}$, we also consider the role of the jet beaming angle, $\theta_{j}$, plays in our understanding of higher isotropic energies for radio loud lGRBs.  A larger $E_{iso}$ may not necessarily indicate a higher energy budget but, instead, a narrower beaming angle. A narrower jet may indeed occur in systems with more angular momentum (Hurtado et al. in prep) and in systems with denser stellar envelopes \cite[i.e. more massive, lower metallicity stars, as shown in][]{LR20}.   
 However, in our data sample, we do {\em not} find a difference between the distribution of the beaming angles between the radio loud and quiet samples \cite[the measured beaming angles of GRB jets, when available, can be found in the metadata table compiled by][]{Wang20}, although we note the numbers of lGRBs with jet opening angle measurements in our samples are small and accounting for the complicated selection effects when measuring beaming angle is difficult \cite[a more detailed description of can be found in][]{LR20}.  Nonetheless, the data currently suggest that the higher isotropic energy measurement in radio loud lGRBs is not a beaming angle effect, but an indication of a truly higher energy/power budget in these systems. 

\subsection{Circumburst Media and Radio Light Curves}


\begin{figure*}
    \centering
    \includegraphics[width=0.46\textwidth]{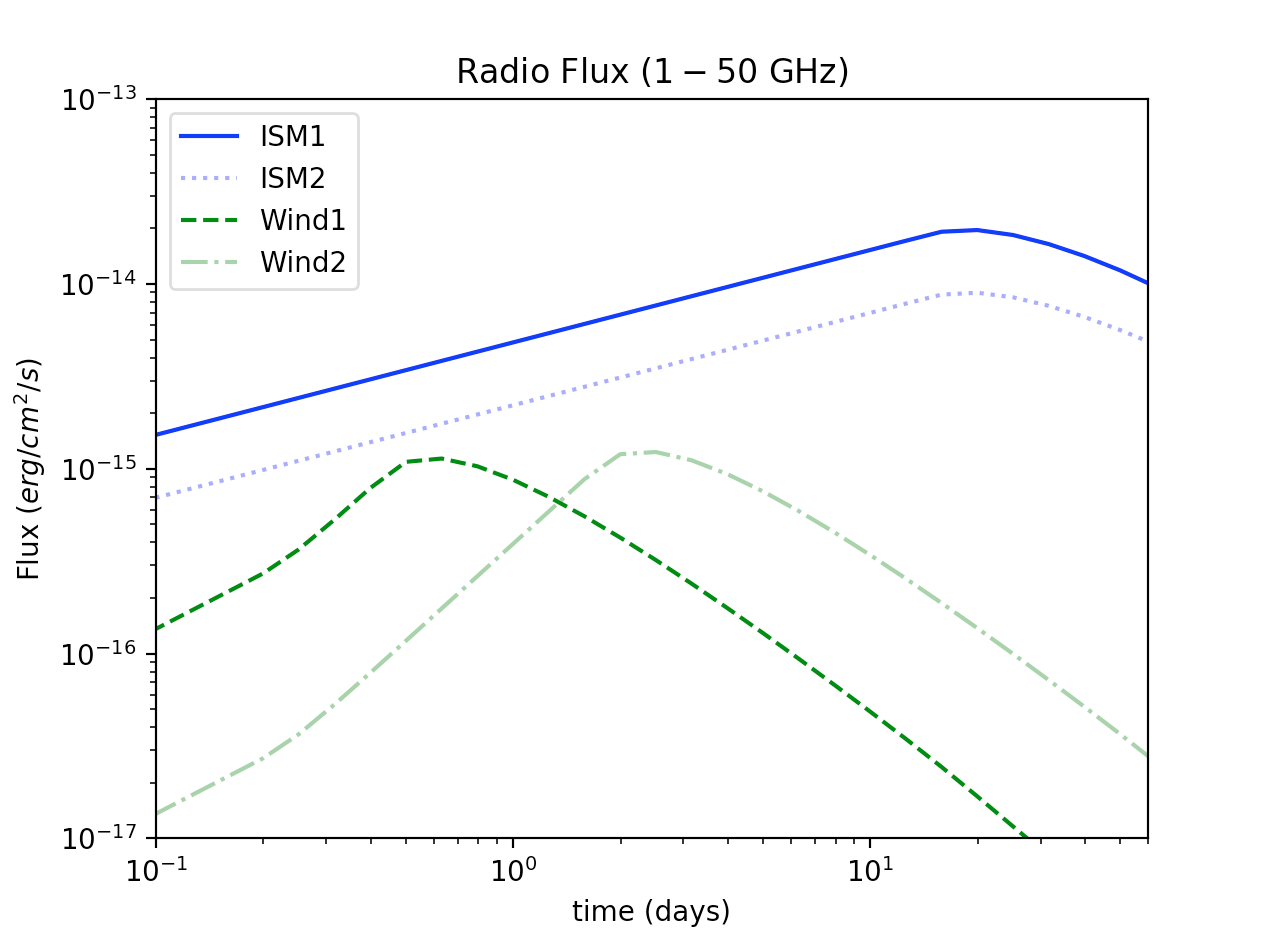} \includegraphics[width=0.46\textwidth]{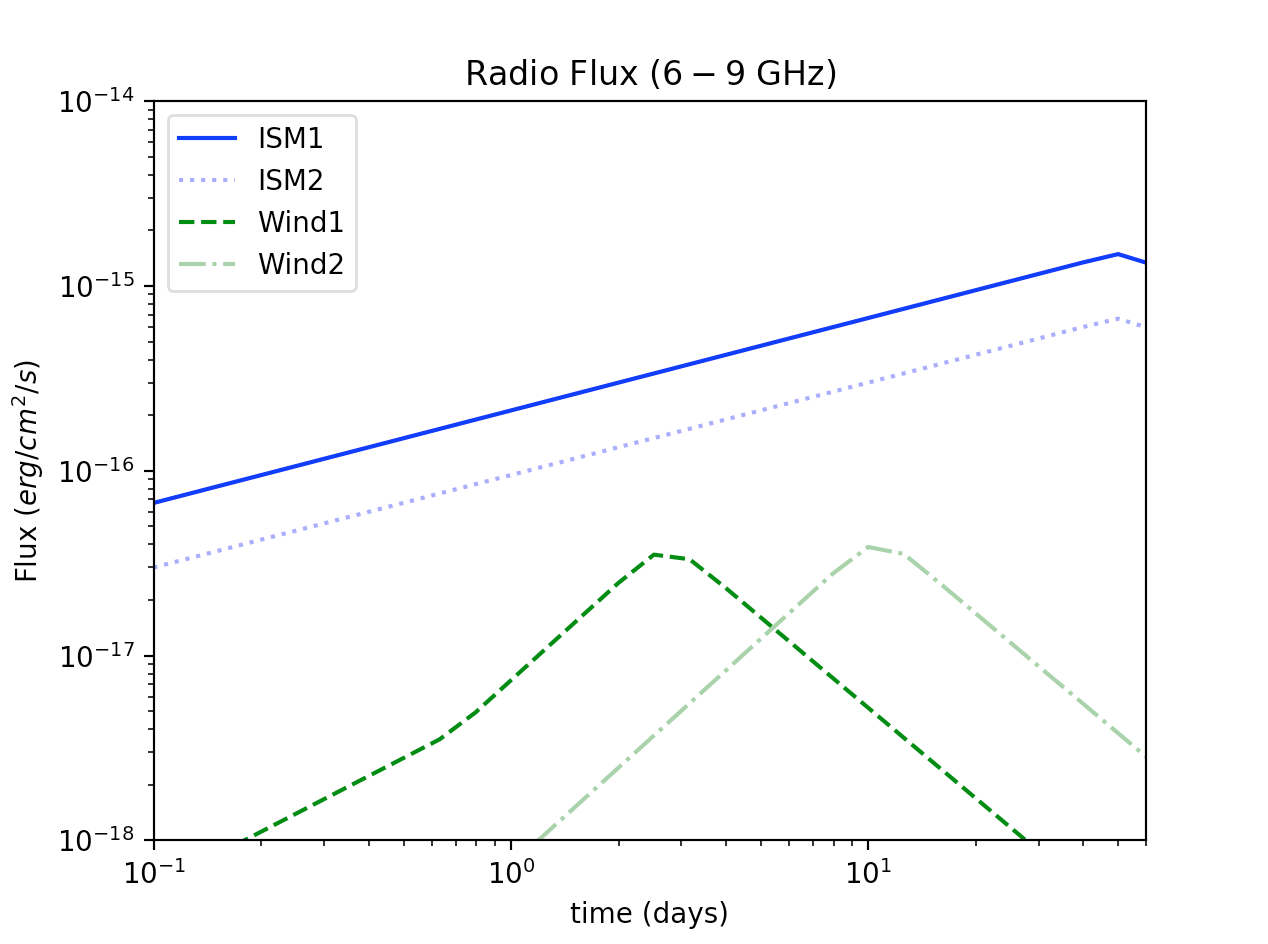} \\
    \includegraphics[width=0.46\textwidth]{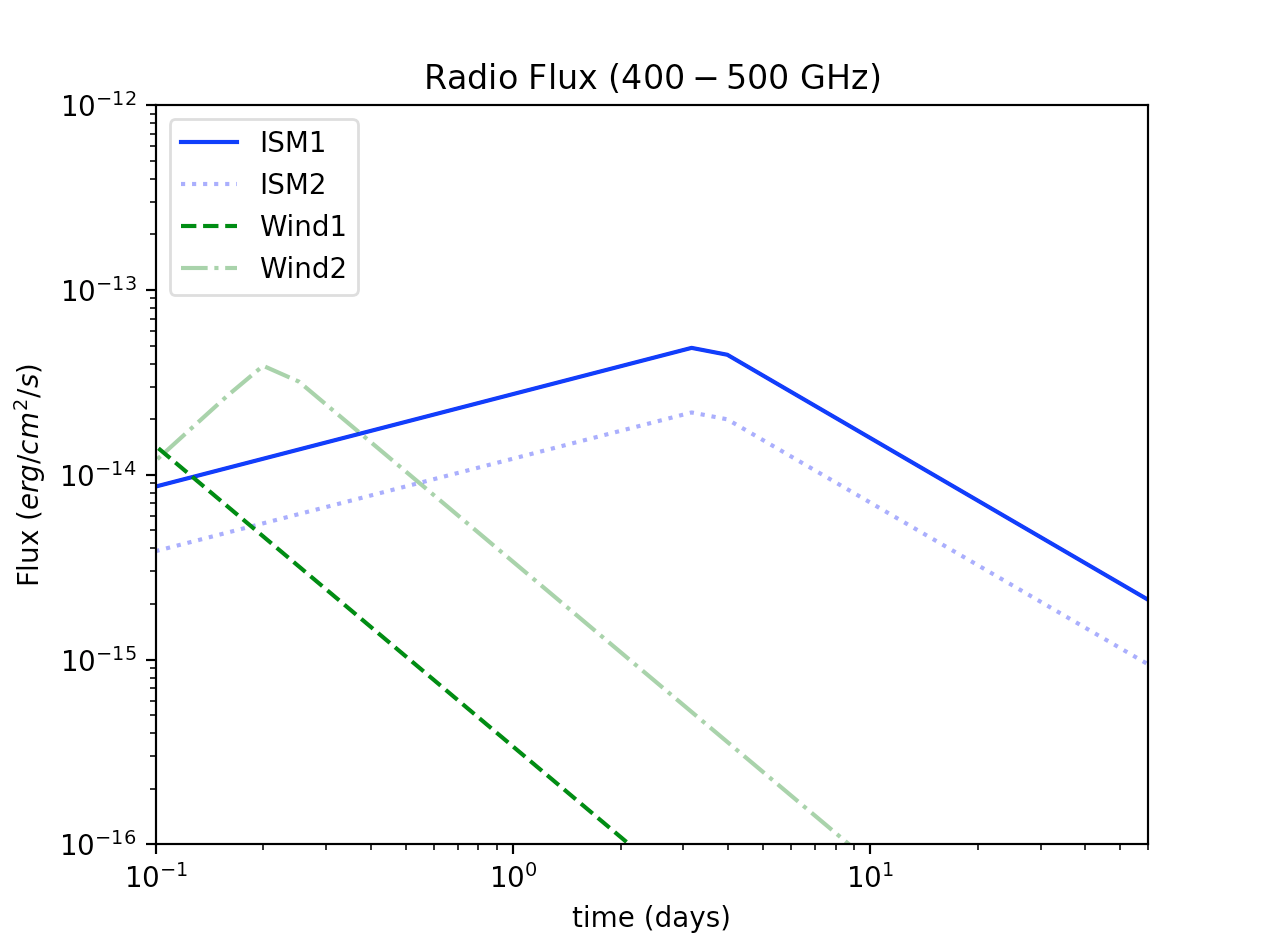} \includegraphics[width=0.46\textwidth]{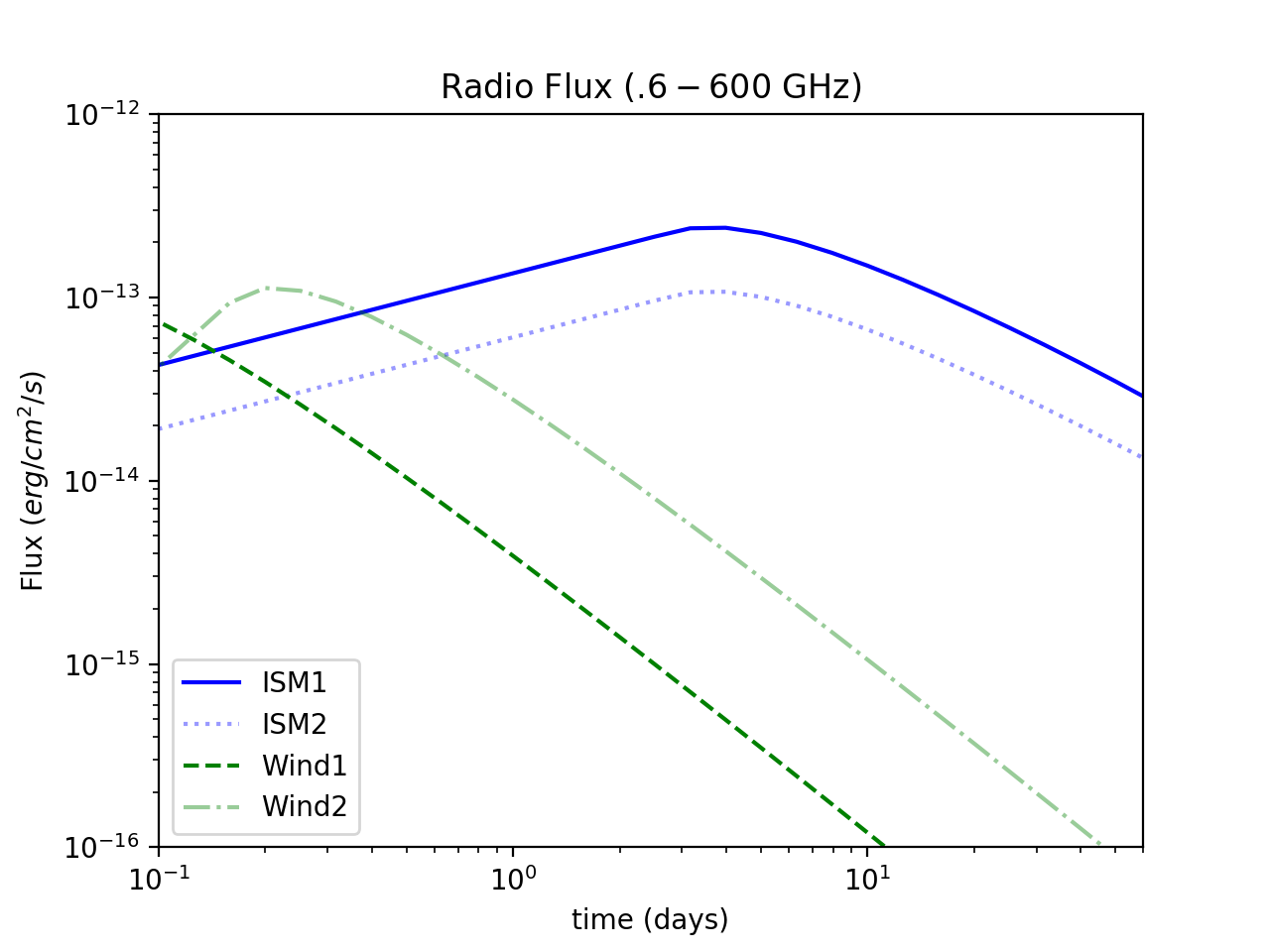}
    \caption{Representative radio light curves in different radio frequency ranges, assuming synchrotron radiation, for jets traversing wind (green dash and dash-dot lines) and ISM-like (blue solid and dotted lines) media. In both cases (ISM and wind) we assume an isotropic energy of $10^{52}$ ergs, an electron energy distribution power law index $p=2.5$, a magnetic energy fraction of $\epsilon_{B} = 0.01$, an electron energy fraction of $\epsilon_{e} = 0.1$, and a redshift $z=1$.  For the ISM curves, we assume a constant density of $n_{o} = 50 \ cm^{-3}$ (solid blue line) and $n_{o}=10 \ cm^{-3}$ (dotted blue line).  For the wind curves, we assume a wind normalization $A_{*}$ (where $\rho = A_{*}r^{-2}$) of $A_{*} = 5 \times 10^{12} g/cm$ (green dashed-dot curve) and $A_{*} = 5 \times 10^{11} g/cm$ (green dash curve). }
    \label{fig:my_label}
\end{figure*}

\begin{figure*}
    \centering
    \includegraphics[width=0.46\textwidth]{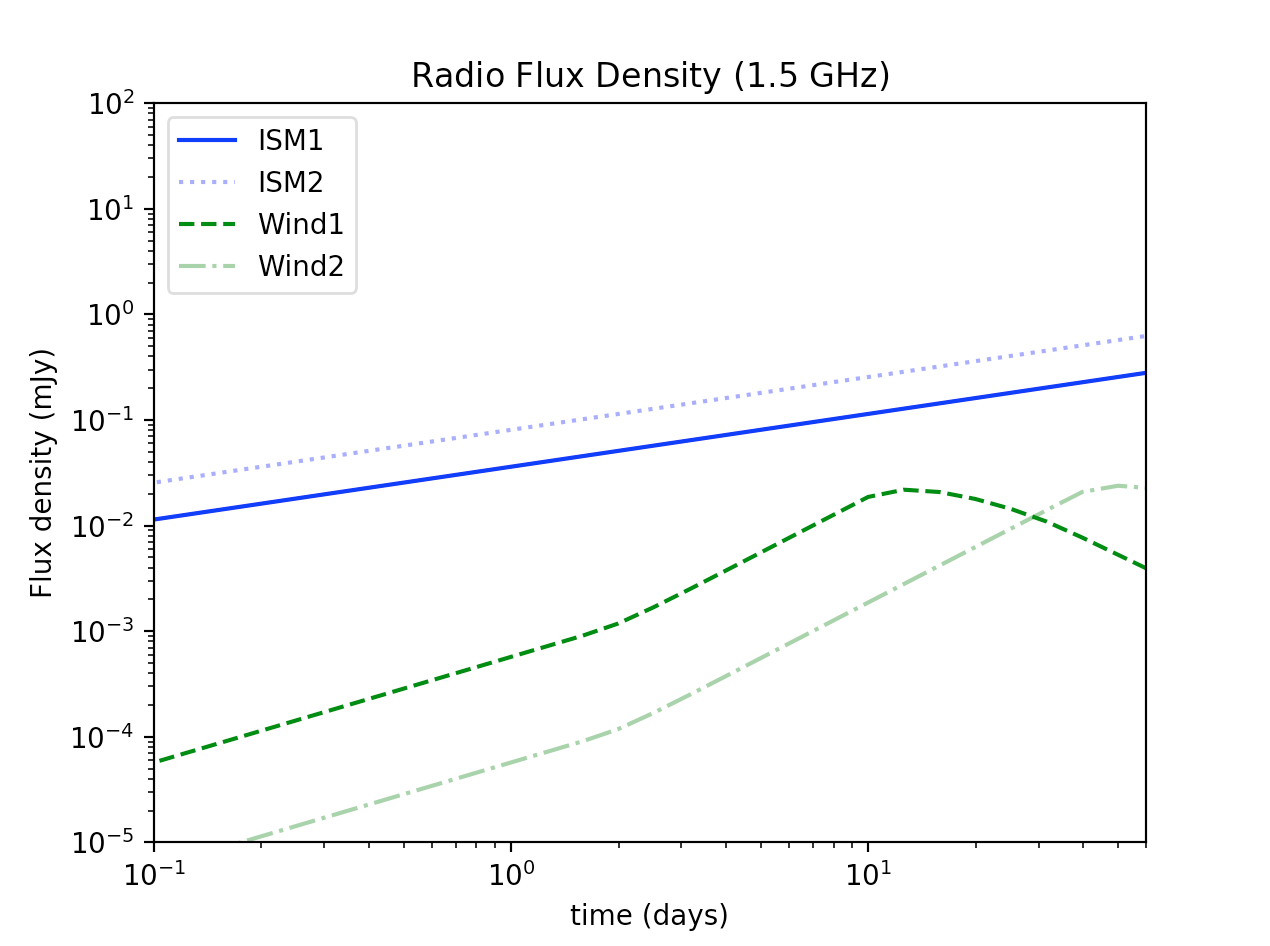} \includegraphics[width=0.46\textwidth]{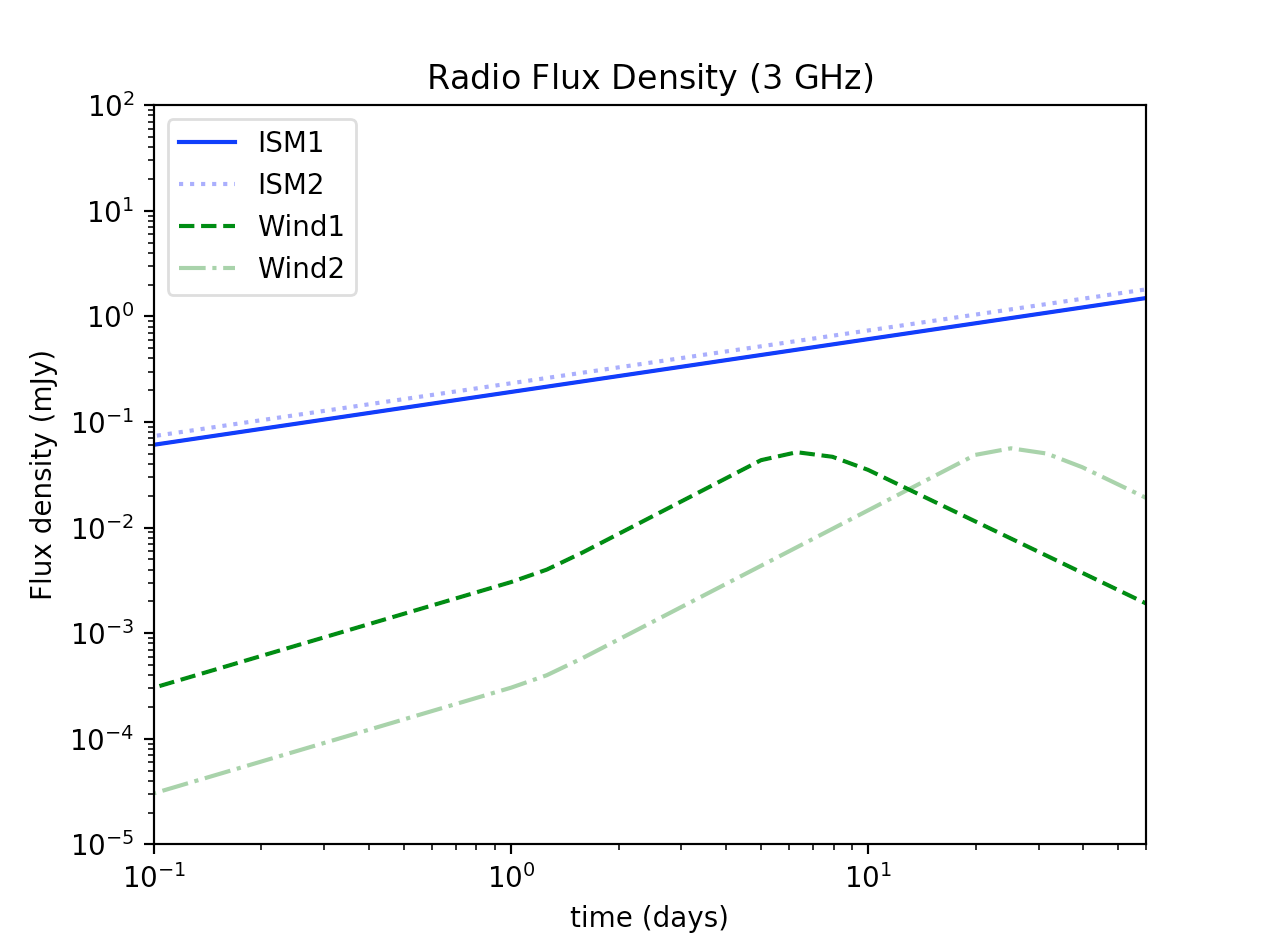} \\
    \includegraphics[width=0.46\textwidth]{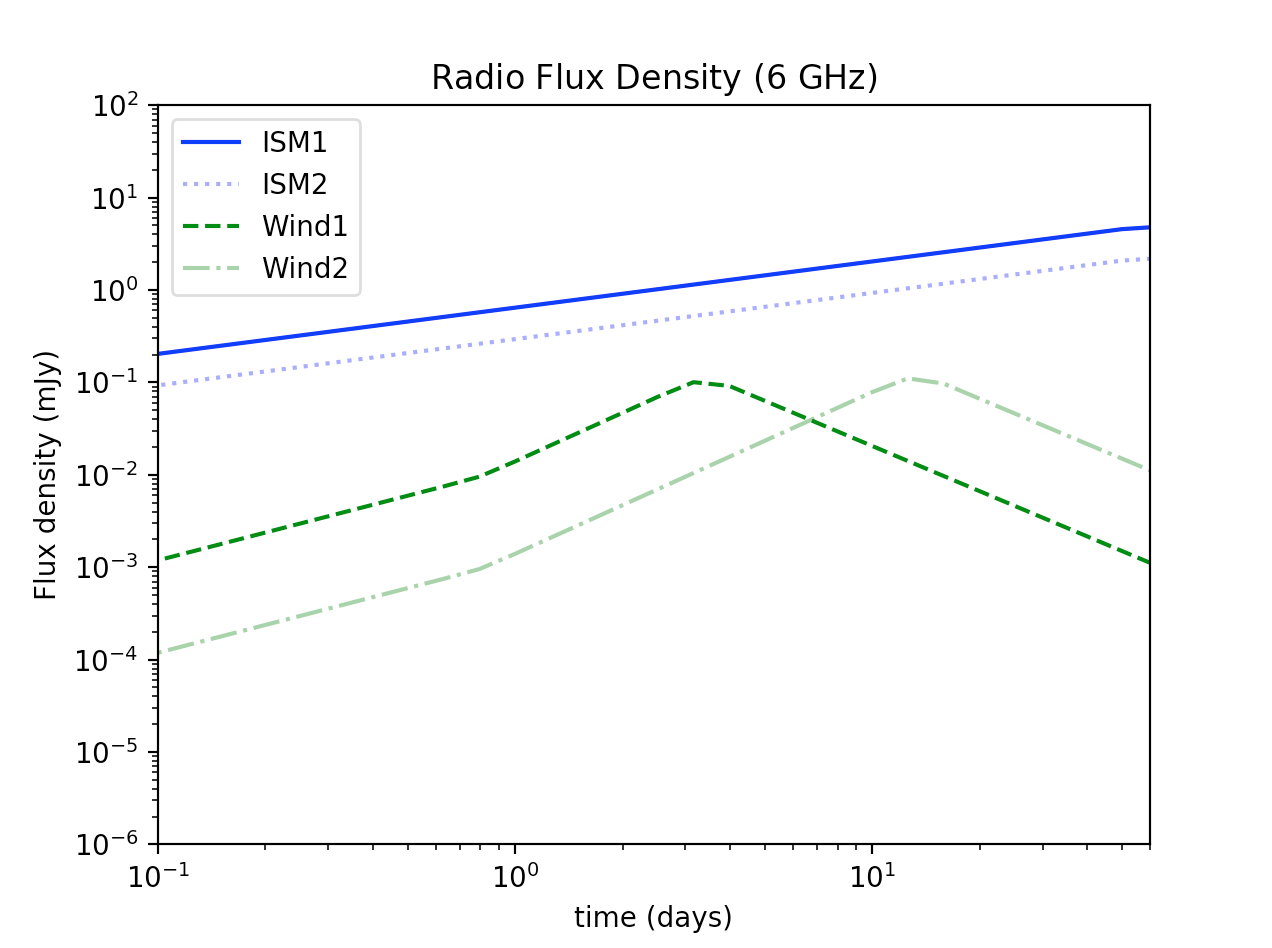} \includegraphics[width=0.46\textwidth]{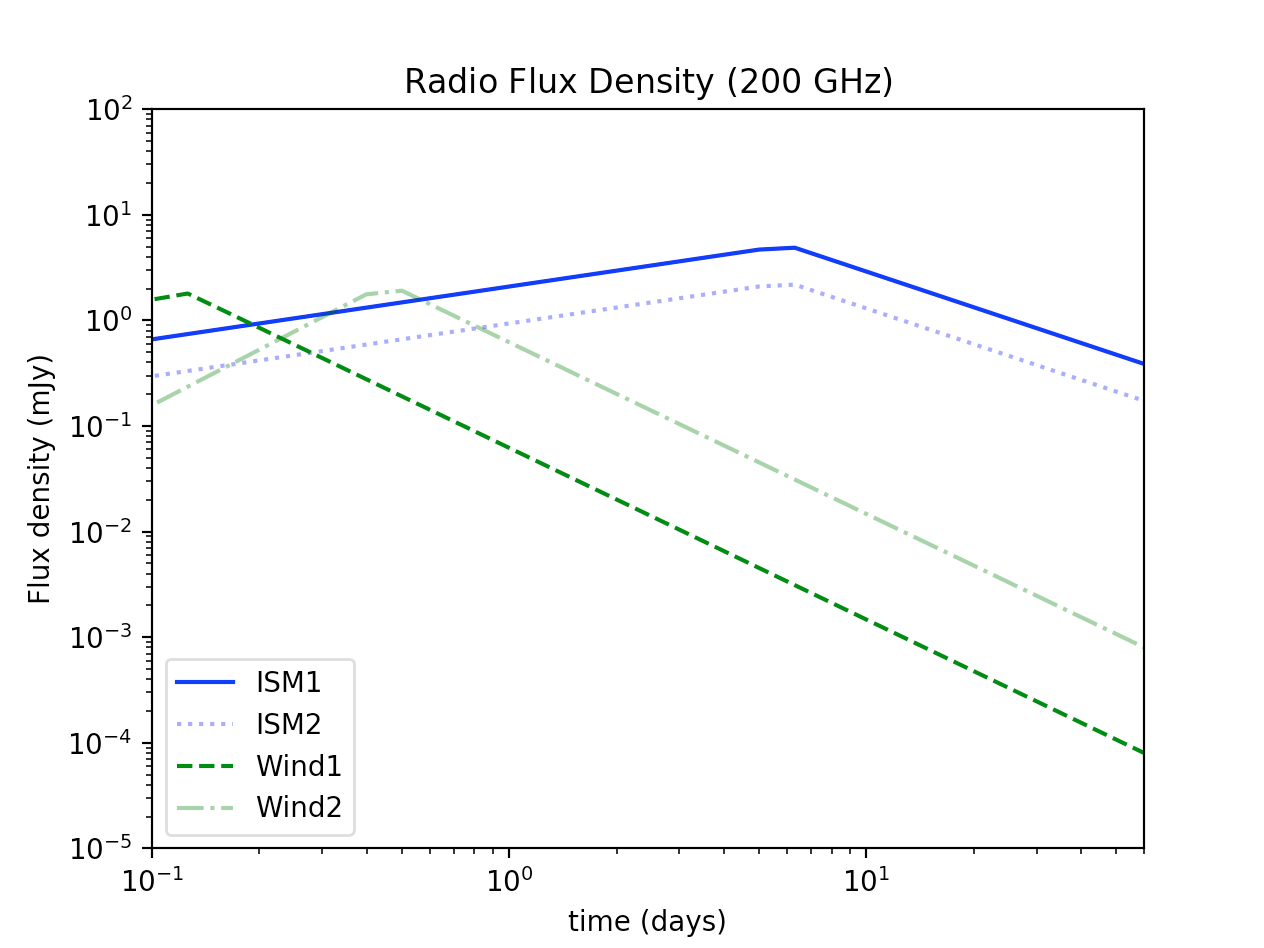}
    \caption{Flux density (mJy) as a function of time for typical radio observational frequencies, assuming synchrotron radiation, for jets traversing wind (green dash and dash-dot lines) and ISM-like (blue solid and dotted lines) media. In both cases (ISM and wind) we assume an isotropic energy of $10^{52}$ ergs, an electron energy distribution power law index $p=2.5$, a magnetic energy fraction of $\epsilon_{B} = 0.01$, an electron energy fraction of $\epsilon_{e} = 0.1$, and a redshift $z=1$.  For the ISM curves, we assume a constant density of $n_{o} = 50 \ cm^{-3}$ (solid blue line) and $n_{o}=10 \ cm^{-3}$ (dotted blue line).  For the wind curves, we assume a wind normalization $A_{*}$ (where $\rho = A_{*}r^{-2}$) of $A_{*} = 5 \times 10^{12} g/cm$ (green dashed-dot curve) and $A_{*} = 5 \times 10^{11} g/cm$ (green dash curve). }
    \label{fig:my_label}
\end{figure*}

Interacting binaries can go through a phase in their evolution that can create a complicated, enhanced environment around the system \citep{Bate00, Olof15, Lines15, Lap20, Schr21}, compared to single star systems in which a massive star's wind is expected to produce a roughly $1/r^{2}$ density profile around the star \citep{CH00}. \cite{Cast21} show, both analytically and with the code {\em Guacho}, that extended, dense shells can form as a result of the interacting winds around two giants in a binary system.  Similarly, \cite{Schr21} show how winds from a star in an interacting binary can produce a a dense, extensive gas structure around the system, to large radii. \cite{Bate00} show that the density profiles in interacting binaries can take on a wide range of profiles, lying between single star wind profiles and uniform density profiles, but more often than not resembling the latter (we note, however, they are examining the circumbinary medium at the formation of the binary system). \\

Further detailed modelling of density profiles around stars in interacting binary systems with at least one massive star companion (i.e. relevant to GRB progenitors) is needed, as there are many system variables to examine \citep{Olof15}. However, it is clear that deviations from a simple $1/r^{2}$ density profile are not unexpected in interacting binaries, and indeed a profile that may appear like a relatively dense uniform density profile (in terms of inferring this profile from the observed GRB emission) is reasonable (see the papers by, e.g., \cite{NG07} and \cite{Gat13}, who show that modest circumburst density fluctuations will not create large fluctuations in the observed GRB light curve; this can lead to spectral and light curve fits that give an apparent constant density profile for the circumburst medium). \\




Now, consider a gamma-ray burst occurring in a constant density medium vs. a wind-like medium.  The {\em peak} of the radio spectrum at the characteristic synchrotron frequencies depends of course on the circumburst density profile.  The peak flux at the synchrotron self-absorption frequency $\nu_{sa}$ (the frequency at which the system is optically thick to synchrotron photons) and the so-called the minimum electron frequency $\nu_{m}$(the frequency that corresponds to the peak of the electron energy distribution) are proportional to the values of the density (or wind normalization density factor) and time as follows \citep{Pach70,SPN98,GS02}: \\

\begin{equation}
\Bigg\{
\begin{aligned}  
& F_{p,\nu_{sa}, ISM}  \propto n^{7/10} t_{day}^{1/2} \\
& F_{p,\nu_{m}, ISM} \propto n^{1/2}t_{day}^{0}
\end{aligned}
\end{equation} 

\begin{equation}
\Bigg\{
\begin{aligned}  
& F_{p,\nu_{sa}, wind}  \propto A_{*}^{7/5}t_{day}^{-1/2} \\
& F_{p,\nu_{m}, wind} \propto A_{*}t_{day}^{-1/2}
\end{aligned}
\end{equation}

\noindent where $n_{o}$ is a constant density in units of $cm^{-3}$ and $A_{*}$ is the normalization in units of $g/cm$ for a wind like density profile, $\rho(r) = A_{*}r^{-2}$. \\

These proportionalities illustrate the more rapid decay of the peak of the lGRB radio spectrum in a wind-like medium, compared to an ISM-like medium.  However, to get a better sense of the relative strength (and detectability) of radio emission from a wind or ISM-like medium, we need to compute the full light curves over the relevant observational frequency range.  In Figure 3, we do just that: we have computed the flux in the radio band as a function of time, over a range of frequencies (that span ranges over which most afterglow searches are performed) for both ISM and wind-like media. We keep careful track of the relative values of the synchrotron self-absorption frequency and minimum electron frequency, throughout the evolution of the light curve \cite[see Figure 1 and Table 2 of][]{GS02}. \\

We use the following expressions for the approximate synchrotron spectrum across different frequency ranges in the ISM and wind cases \citep{GS02}: 
\begin{equation}
\begin{aligned}
    A_{ISM} = & 1.18 \times 10^{8} mJy \ (4.59 - p) (1+z)^{9/4} \\
    & \epsilon_{B}^{-1/4} n^{-1/2} E_{52}^{1/4} t_{days}^{5/4}d_{L, 28}^{-2} \nu_{14}^{5/2} \\  
\end{aligned}
\end{equation}

\begin{equation}
    \begin{aligned}
    B_{ISM} = & 4.20 \times 10^{9} mJy \ \frac{3p+2}{3p-1} (1+z)^{5/2} \\
    & \bar{\epsilon_{e}} n^{-1/2} E_{52}^{1/2} t_{days}^{1/2}d_{L, 28}^{-2} \nu_{14}^{2} \\  
\end{aligned}
\end{equation}

\begin{equation}
    \begin{aligned}
    D_{ISM} = & 27.9 mJy \ \frac{p-1}{3p-1} (1+z)^{5/6} \\
    & \bar{\epsilon_{e}}^{-2/3} \epsilon_{B}^{1/3} n^{-1/2} E_{52}^{5/6} t_{days}^{1/2}d_{L, 28}^{-2} \nu_{14}^{1/3} \\  
\end{aligned}
\end{equation}

\begin{equation}
\begin{aligned}
   G_{ISM} = & 0.461 mJy \ (p-.04)e^{2.53p} (1+z)^{(3+p)/4} \\
    & \bar{\epsilon_{e}}^{p-1} \epsilon_{B}^{(1+p)/4} n^{-1/2} E_{52}^{(3+p)/4} t_{days}^{3(1-p)/4} \\
    & d_{L, 28}^{-2} \nu_{14}^{(1-p)/2} \\ 
\end{aligned}
\end{equation} \\

\noindent where $z$ is redshift, $p$ is the index of a power-law energy distribution for the radiating electrons, $n$ is density in units of $cm^{-3}$, $\epsilon_{B}$ is the fraction of energy in the magnetic field, $\bar{\epsilon_{e}}$ is the fraction of energy in the electrons, $\epsilon_{e}$, times the factor $(p-2)/(p-1)$, $E_{52}$ is the energy in units of $10^{52}$ ergs, $t_{day}$ is observation time in units of days, $d_{L, 28}$ is the luminosity distance in units of $10^{28}cm$, $\nu_{14}$ is the frequency in units of $10^{14}Hz$. \\ \\

In the wind circumburst medium case we have:
\begin{equation}
    \begin{aligned}
    A_{wind} = & 2.96 \times 10^{7} mJy \ (4.59 - p) (1+z)^{7/4} \\
    & \epsilon_{B}^{-1/4} A_{*}^{-1} E_{52}^{3/4} t_{days}^{7/4}d_{L, 28}^{-2} \nu_{14}^{5/2} \\  
\end{aligned}
\end{equation}

\begin{equation}
    \begin{aligned}
    B_{wind} = & 1.33 \times 10^{9} mJy \ \frac{3p+2}{3p-1} (1+z)^{2} \\
    & \bar{\epsilon_{e}} A_{*}^{-1} E_{52} t_{days}d_{L, 28}^{-2} \nu_{14}^{2} \\  
\end{aligned}
\end{equation}

\begin{equation}
    \begin{aligned}
    D_{wind} = & 211 mJy \ \frac{p-1}{3p-1} (1+z)^{4/3} \\
    & \bar{\epsilon_{e}}^{-2/3} \epsilon_{B}^{1/3} A_{*} E_{52}^{1/3} t_{days}^{0}d_{L, 28}^{-2} \nu_{14}^{1/3} \\  
\end{aligned}
\end{equation}

\begin{equation}
   \begin{aligned}
   G_{wind} = & 3.82 mJy \ (p-.18)e^{2.54p} (1+z)^{(5+p)/4} \\
    & \bar{\epsilon_{e}}^{p-1} \epsilon_{B}^{(1+p)/4} A_{*} E_{52}^{(1+p)/4} t_{days}^{(1-3p)/4} \\
    & d_{L, 28}^{-2} \nu_{14}^{(1-p)/2} \\ 
\end{aligned}
\end{equation} \\

\noindent where, again, $A_{*}$ is the wind density profile normalization factor in units of $g/cm$.\\ \\

To compute the radio light curve , then, our prescription is as follows: In the case when the self-absorption frequency is greater than the minimum electron frequency, the flux between observed frequencies $\nu_{1}$ and $\nu_{2}$ for an ISM/wind medium is given by:
\begin{equation}
    F = \int_{\nu_{1}}^{\nu_{sa}} B_{k} d\nu + \int_{\nu_{sa}}^{\nu_{m}} D_{k} d\nu + \int_{\nu_{m}}^{\nu_{2}} G_{k} d\nu 
\end{equation}

\noindent where $k$ indicates whether we are in an ISM ($k=0$) or wind ($k=2$) medium (this is based on how density varies with radius, $\rho \propto r^{-k}$), in the respective media.\\

When the self-absorption frequency $\nu_{sa}$ is less than the minimum electron frequency, we have:
 \begin{equation}
    F = \int_{\nu_{1}}^{\nu_{sa}} B_{k} d\nu + \int_{\nu_{sa}}^{\nu_{m}} A_{k} d\nu + \int_{\nu_{m}}^{\nu_{2}} G_{k} d\nu 
\end{equation}.
\noindent To compute the flux densities (flux per unit frequency) shown in Figure 4, we omit the integration and evaluate the $A, B, C$, and $D$ expressions at a particular relevant observational frequency.

The above expressions assume the physical break frequencies are within the observed band; i.e. $\nu_{1}< \nu_{sa} < \nu_{m} < \nu_{2}$ for equation 21 and $\nu_{1}< \nu_{m} < \nu_{sa} < \nu_{2}$ for equation 22.  Of course if the physical break frequencies fall outside the observed band, we adjust the limits of our integral accordingly. \\

In Figure 3, we show light curves for systems with an isotropic energy of $10^{52}$ ergs, an electron energy distribution power law index $p=2.5$, a magnetic energy fraction of $\epsilon_{B} = 0.01$, an electron energy fraction of $\epsilon_{e} = 0.1$, and a redshift $z=1$.  For the ISM curves, we assume a constant density of $n_{o}=10 \ cm^{-3}$ (dotted blue line) and $n_{o} = 50 \ cm^{-3}$ (solid blue line).  For the wind curves, we assume a wind normalization $A_{*} = 5 \times 10^{12} g/cm$ (green dashed-dot curve) and $A_{*} = 5 \times 10^{11} g/cm$ (green dash curve). Figure 4 shows the radio flux densities in at specific relevant observational frequencies.  Certainly we have the freedom to adjust the normalization of these curves based on the values of the physical parameters we choose; however, we have chosen relatively conservative and typical values expected for GRBs, based on both theory and fits to their afterglow light curves \citep{vE13,Gomp18,KF21}.\\

It is clear that the constant density medium case, even for relatively modest densities, is brighter and longer-lasting that the wind density case, indicating a higher likelihood of detecting a radio afterglow in such an environment.\footnote{We note  that fits to broadband light curves and spectra give inconclusive results for the density profiles.  \cite{GFP18} fit a sample of lGRBs to both wind and ISM medium density models and did not find that radio loud GRBs were better fit by one density profile over the other. However, because of the simplicity of the density models and the degeneracy of density with other model parameters (as well as small number statistics; there were only 15 GRBs from our radio loud GRBs in their sample), it is difficult to draw definitive conclusions about the external density profiles from these fits. We consider this an open issue to be further explored. } \\

Our main point here is that, for fiducial values of intrinsic GRB properties, we expect brighter radio emission in systems with an extended ISM-like environment, which we suggest is more likely in an interacting binary system compared to a single massive star system.

\section{Rates}
 In our sample of energetic lGRBs with radio follow-up, we find that roughly 60 \% of GRBs appear to be radio loud. Accordingly, from their radio image stacking analysis of lGRBs, \cite{HGM13} suggested that roughly 30 to 40 \% are truly radio quiet. In Figure 5, we plot the radio loud fraction of lGRBs as a function of redshift for five redshift bins.  Although it appears to evolve somewhat (peaking at around $(1+z) \approx 2.5$), we caution that within the large error bars, this ratio is consistent with staying roughly constant over redshift.  We would like to compare this - the fraction of lGRBs that are radio loud - to the binary star fraction (more specifically, the fraction of binary systems that can produce an lGRB).
 
  Unfortunately, this question is plagued with many complicating factors.  First of all, getting an accurate handle on overall binary rates, through both observations and binary population synthesis studies, is notoriously difficult \citep{Han20, DW20, Mazz20}. Furthermore, determining those massive stars in close, interacting binaries that would produce an lGRB has additional complexities associated with understanding the conditions necessary for a successful GRB.  
  
  Several studies \cite[e.g.][]{Sana12,Sana13} have put the binary star fraction at around 60 \%, with some studies \citep{MG09} showing rates closer to 75 \%.   Although this percentage is interestingly similar to what we find for the fraction of radio loud GRBs, we again caution that directly comparing these percentages ignores the details involved in understanding the rates of those binary star systems that could create lGRBs.  Nonetheless, if we can make the simplistic but reasonable assumption that the fraction of single stars that create lGRBs is roughly equivalent to the fraction of binary stars that create lGRBs (more on this below), then, combined with the other factors described in \S 3 for why interacting binary progenitors are consistent with the data from radio loud lGRBS, we find the similarity of the ratios suggestive enough to warrant further investigation of this issue.

Consider the rate of lGRBs ($\dot{dN}/dz)$ as it relates to the star formation rate $\dot{\rho}_{\rm SFR}(z)$ \citep{Kist08,Yuk08,Kist09, LR20}:

\begin{equation}
    (\dot{dN}/dz)=  \dot{\rho}_{\rm SFR}(z)/({\rm f_{beam}(z)})  \left(\frac{dV/dz}{(1+z)} \right) \epsilon(z)
\end{equation}
\noindent where $f_{beam}(z)$ is a factor ($>1$) that accounts for the number of GRBs missed due to GRB jets that are pointed away from our detectors, and $\epsilon(z)$ parameterizes the fraction of stars that make GRBs.  Both of these functions can in principle depend on redshift (\cite{LR20} show that indeed $\rm f_{beam}(z)$ does).  The factor $dV/dz$ is the cosmological volume element given by:
 \begin{equation}
\begin{split}
    dV/dz = & 4 \pi (\frac{c}{H_{o}})^{3} \bigg[\int_{1}^{1+z} \frac{d(1+z)}{\sqrt{\Omega_{\Lambda} + \Omega_{m}(1+z)^{3}}}\bigg]^{2} \\
    & \times \frac{1}{\sqrt{\Omega_{\Lambda} + \Omega_{m}(1+z)^{3}}}
\end{split}
\end{equation}

 \noindent where $\Omega_{m}$, $\Omega_{\Lambda}$ are the matter and cosmological constant density parameters, and the $H_{o}$ is the Hubble constant.

If we can assume that a function $\epsilon(z)$ is a sufficient way to account for how each population (i.e. radio loud or radio quiet) independently traces the star formation rate, and if we assume that the jet beaming angle and its evolution over cosmic time are roughly similar in both radio loud and quiet populations, then the fraction of radio loud lGRBs is the ratio of the integral of $\epsilon(z)$ for the radio bright sample relative to the whole population:
\begin{equation}
    \frac{\int (\dot{dN}/dz_{RL})}{\int (\dot{dN}/dz)} =  \frac{\int \epsilon_{RL}(z)}{\int\epsilon_{RQ}(z) +  \int \epsilon_{RL}(z)} \approx 0.6 \pm 0.2
\end{equation}

The functions $\epsilon(z)$ are as yet undetermined for the progenitor models we suggest here (i.e. collapsars in interacting binaries and single star/non-interacting binary collapsars; again, this function is parameterizing the fundamental, underlying question of which stars produce GRBs and why).  As we described above, to fairly compare the binary star fraction to the radio loud lGRB fraction (or, alternatively, the single star fraction to the radio quiet lGRB fraction), we need to assume that $\epsilon(z)$ is roughly similar for both radio loud and radio quiet progenitors.   \\


\begin{figure}
    \centering
    \includegraphics[width=0.5\textwidth]{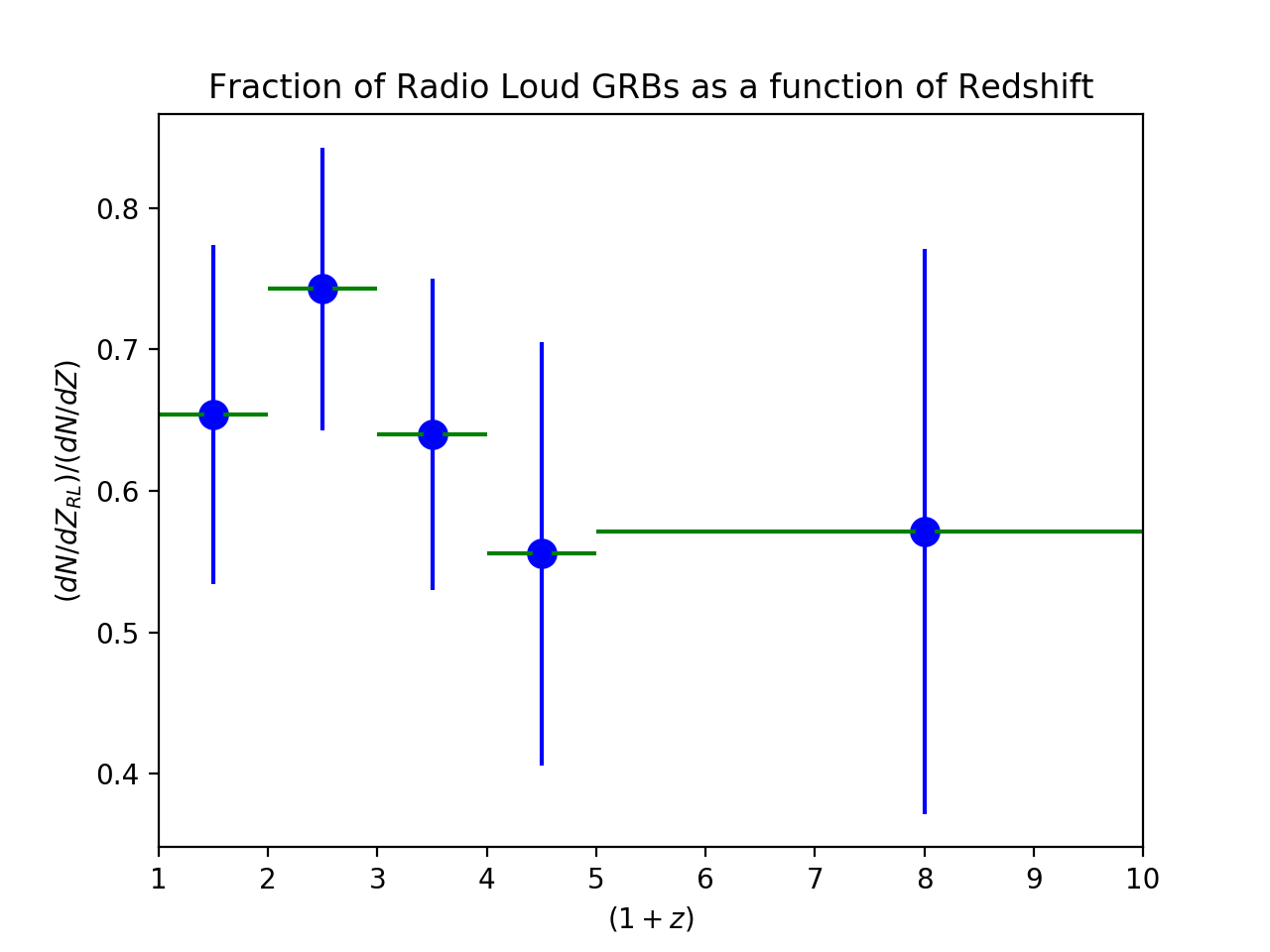}
    \caption{Fraction of radio loud GRBs (relative to the total population), for those with attempted radio follow-up, as a function of redshift. Data are taken from the sample described in \cite{LR19}, who use a subset of the data published in \cite{CF12}.}
    \label{fig:my_label}
\end{figure}

  We again mention the simulations of \cite{Bav21}, who indicate that common envelope (CE) and chemically homogeneous evolution (CHE) interacting binaries (ones that would apply to systems we are discussing here) add up to $= 71 \%$ ($57 \%$ (CE) $+ 14 \%$ (CHE)) of the binary systems they consider (note they argue binary systems can explain the rates of the entire population of lGRBs; this could suggest, as mentioned above, that the radio loud vs radio quiet is more a reflection of the type of binary interaction rather than binary vs. single star systems).
  
  Finally, we comment that metallicity plays an important role in all of the considerations above. It has long been suggested that low metallicity is a necessary condition to produce a successful GRB jet from a collapsing star because it mitigates mass and angular momentum loss \citep{MW99,YL05,HMM05,Yoon06,WH06}, and indeed lGRBs tend to be found in lower metallicity galaxies (for a recent review summarizing these observational results, see \cite{Per16}). Meanwhile, the close binary fraction may be higher for massive and lower metallicity stars \citep{DM91, Bav21, Klen21}. \cite{Mazz20} show, using APOGEE data, an anti-correlation between the close binary fraction and [Fe/H] ratio, highlighting the necessity of low metallicity in producing these systems (see their Figure 5; similarly \cite{Hwang21} present recent results showing the wide binary fraction is positively correlated with metallicity).  On the other hand, \cite{Liu21} show that low metallicity (specifically Pop III) binary systems are born wider. However, \cite{Det08} show that massive star winds widen binary systems so low metallicity is needed to achieve close binary systems.   \cite{Schr21} show that either widen or tighten with metallicity depending on the velocity of the wind (relative to the orbital velocity) and the mass ratio of the binary. The detailed connection between metallicity, GRBs, and interacting binary systems is beyond the scope of this paper but is important factor in future work investigating these systems.

 \section{Conclusions}
The radio emission from a long GRB is a reflection of a GRB jet's interaction with its surrounding external medium, while the duration of the prompt gamma-ray emission comes from internal dissipation processes in the jet and reflect (along with the energy budget) the system's inner engine.  That lGRBs with radio afterglows appear have significantly longer gamma-ray durations and higher isotropic energies than those with no radio afterglows, suggests that radio loud and radio quiet lGRBs may originate from different progenitor systems.  \\

We have argued broadly in this paper that these observational results can be explained if radio loud GRBs emanate from interacting binary systems, while radio quiet GRBs emerge from the collapse of a single star (or a massive star collapsing in a non-interacting binary, where we define interacting binary as one where the companion has significant influence on the primary star in terms of mass and angular momentum transfer).  Our main arguments are:

\begin{itemize}
    \item A massive star in a closely interacting binary can have significant angular momentum, due to spin up from its companion. This can produce both a more massive disk and lower accretion rate in the system, leading to a longer duration gamma-ray burst.
    \item A massive star in a closely interacting binary may gain mass from its companion which can lead to a higher isotropic energy when the star collapses. The higher isotropic energies seen in the radio loud sample of lGRBs may also be a reflection of the system's higher angular momentum and its ability to provide more power to the jet, given a Blandford-Znajek scenario for the jet launch mechanism.   
    \item Radio emission is brighter in an extended, uniform medium. Closely interacting binary systems have the ability to create a more extended, dense circumbinary medium that may lead to brighter, longer-lived radio emission compared to a single star systems with a wind-like medium.
    \item Although we acknowledge the complicating factors that prohibit a direct comparison, the ratio of  binary systems to single star systems aligns with the ratio of radio loud to radio dark GRBs.  
\end{itemize}

 Additional clues, including the presence or absence of extended very high energy emission (shown to be more likely in extended constant density media and not wind-like media) as well as correlations specific to either one of the radio loud or quiet populations, such as the duration redshift anti-correlation shown in Figure 2, can further test our hypothesis. Correlations of the rates of radio loud lGRBs with that of high mass X-ray binaries may also offer additional clues (Upton-Sanderbeck, private communication).  Nonetheless, as we accumulate and analyze more lGRB afterglow data, it is becoming clear that the radio emission in particular shows unusual behavior and has challenged our standard model of lGRBs. For example, recently, \cite{KF21,Mar21} and Dvornak et al (in prep) show that the standard synchrotron shock emission models do not fit the radio light curves well.  \cite{Lev21} show that the break times and plateau-like behavior seen in radio afterglow light curves are unique compared to optical and X-ray plateaus.

As such, radio observations of lGRBs may offer us a unique and deeper insight into the underlying physics of these objects. Continued radio follow-up, more observational and theoretical work on the formation of binary and single stars, as well as a deeper investigation into the conditions necessary to produce a relativistic GRB jet, will allow us to better test the idea presented in this paper and further understand the fascinating processes behind lGRBs and their role in our universe.

\section{Acknowledgements}
We are very grateful to the referee for thoughtful and helpful commments that improved this manuscript.  We thank Olivia Cantrell Rodriguez, Jarrett Johnson, Angana Chakraborty, Ken Luu, Roseanne Cheng, Maria Dainotti, Valeria Hurtado, Celia Tandon, Aycin Aykutalp, Phoebe Upton-Sanderbeck, Delina Levine and Kevin Zvonarek for lively and interesting discussions related to this work.  Los Alamos National Laboratory is operated by Triad National Security, LLC, for the National Nuclear Security Administration of U.S. Department of Energy (Contract No. 89233218CNA000001).  LA-UR-22-20471
 
\bibliography{refs}

\begin{thebibliography}{}
\expandafter\ifx\csname natexlab\endcsname\relax\def\natexlab#1{#1}\fi
\providecommand{\url}[1]{\href{#1}{#1}}
\providecommand{\dodoi}[1]{doi:~\href{http://doi.org/#1}{\nolinkurl{#1}}}
\providecommand{\doeprint}[1]{\href{http://ascl.net/#1}{\nolinkurl{http://ascl.net/#1}}}
\providecommand{\doarXiv}[1]{\href{https://arxiv.org/abs/#1}{\nolinkurl{https://arxiv.org/abs/#1}}}

\bibitem[{{Abbott} {et~al.}(2017){Abbott}, {Abbott}, {Abbott}, {Acernese},
  {Ackley}, {Adams}, {Adams}, {Addesso}, {Adhikari}, {Adya}, \& et~al.}]{Ab17}
{Abbott}, B.~P., {Abbott}, R., {Abbott}, T.~D., {et~al.} 2017, Physical Review
  Letters, 119, 161101, \dodoi{10.1103/PhysRevLett.119.161101}

\bibitem[{{Amati} {et~al.}(2002){Amati}, {Frontera}, {Tavani}, {in't Zand},
  {Antonelli}, {Costa}, {Feroci}, {Guidorzi}, {Heise}, {Masetti}, {Montanari},
  {Nicastro}, {Palazzi}, {Pian}, {Piro}, \& {Soffitta}}]{AM02}
{Amati}, L., {Frontera}, F., {Tavani}, M., {et~al.} 2002, \aap, 390, 81,
  \dodoi{10.1051/0004-6361:20020722}

\bibitem[{{Barkov} \& {Komissarov}(2008)}]{BK08}
{Barkov}, M.~V., \& {Komissarov}, S.~S. 2008, International Journal of Modern
  Physics D, 17, 1669, \dodoi{10.1142/S0218271808013285}

\bibitem[{{Barkov} \& {Komissarov}(2010)}]{BK10}
---. 2010, \mnras, 401, 1644, \dodoi{10.1111/j.1365-2966.2009.15792.x}

\bibitem[{{Bate}(2000)}]{Bate00}
{Bate}, M.~R. 2000, \mnras, 314, 33, \dodoi{10.1046/j.1365-8711.2000.03333.x}

\bibitem[{{Bavera} {et~al.}(2021){Bavera}, {Fragos}, {Zapartas},
  {Ramirez-Ruiz}, {Marchant}, {Kelley}, {Zevin}, {Andrews}, {Coughlin},
  {Dotter}, {Kovlakas}, {Misra}, {Serra-Perez}, {Qin}, {Rocha},
  {Rom{\'a}n-Garza}, {Tran}, \& {Xing}}]{Bav21}
{Bavera}, S.~S., {Fragos}, T., {Zapartas}, E., {et~al.} 2021, arXiv e-prints,
  arXiv:2106.15841.
\newblock \doarXiv{2106.15841}

\bibitem[{{Beniamini} {et~al.}(2020){Beniamini}, {Granot}, \& {Gill}}]{BGG20}
{Beniamini}, P., {Granot}, J., \& {Gill}, R. 2020, \mnras, 493, 3521,
  \dodoi{10.1093/mnras/staa538}

\bibitem[{{Blandford} \& {Znajek}(1977)}]{BZ77}
{Blandford}, R.~D., \& {Znajek}, R.~L. 1977, \mnras, 179, 433,
  \dodoi{10.1093/mnras/179.3.433}

\bibitem[{{Bloom} {et~al.}(2002){Bloom}, {Kulkarni}, \& {Djorgovski}}]{BKD02}
{Bloom}, J.~S., {Kulkarni}, S.~R., \& {Djorgovski}, S.~G. 2002, \aj, 123, 1111,
  \dodoi{10.1086/338893}

\bibitem[{{Castellanos-Ram{\'\i}rez} {et~al.}(2021){Castellanos-Ram{\'\i}rez},
  {Rodr{\'\i}guez-Gonz{\'a}lez}, {Meliani}, {Rivera-Ortiz}, {Raga}, \&
  {Cant{\'o}}}]{Cast21}
{Castellanos-Ram{\'\i}rez}, A., {Rodr{\'\i}guez-Gonz{\'a}lez}, A., {Meliani},
  Z., {et~al.} 2021, \mnras, \dodoi{10.1093/mnras/stab2373}

\bibitem[{{Chandra} \& {Frail}(2012)}]{CF12}
{Chandra}, P., \& {Frail}, D.~A. 2012, \apj, 746, 156,
  \dodoi{10.1088/0004-637X/746/2/156}

\bibitem[{Chevalier \& Li(2000)}]{CH00}
Chevalier, R.~A., \& Li, Z.-Y. 2000, The Astrophysical Journal, 536, 195

\bibitem[{{Chrimes} {et~al.}(2020){Chrimes}, {Stanway}, \& {Eldridge}}]{CSE20}
{Chrimes}, A.~A., {Stanway}, E.~R., \& {Eldridge}, J.~J. 2020, \mnras, 491,
  3479, \dodoi{10.1093/mnras/stz3246}

\bibitem[{{Dainotti}(2019)}]{Dai19}
{Dainotti}, M. 2019, {Gamma-ray Burst Correlations; Current status and open
  questions}, \dodoi{10.1088/2053-2563/aae15c}

\bibitem[{{Dainotti} {et~al.}(2021){Dainotti}, {Levine}, {Fraija}, \&
  {Chandra}}]{Dain21}
{Dainotti}, M., {Levine}, D., {Fraija}, N., \& {Chandra}, P. 2021, Galaxies, 9,
  95, \dodoi{10.3390/galaxies9040095}

\bibitem[{{Dainotti} \& {Amati}(2018)}]{DA18}
{Dainotti}, M.~G., \& {Amati}, L. 2018, \pasp, 130, 051001,
  \dodoi{10.1088/1538-3873/aaa8d7}

\bibitem[{{Dainotti} \& {Del Vecchio}(2017)}]{Dai17}
{Dainotti}, M.~G., \& {Del Vecchio}, R. 2017, \nar, 77, 23,
  \dodoi{10.1016/j.newar.2017.04.001}

\bibitem[{{Dainotti} {et~al.}(2018){Dainotti}, {Del Vecchio}, \&
  {Tarnopolski}}]{Dai18}
{Dainotti}, M.~G., {Del Vecchio}, R., \& {Tarnopolski}, M. 2018, Advances in
  Astronomy, 2018, 4969503, \dodoi{10.1155/2018/4969503}

\bibitem[{{Detmers} {et~al.}(2008){Detmers}, {Langer}, {Podsiadlowski}, \&
  {Izzard}}]{Det08}
{Detmers}, R.~G., {Langer}, N., {Podsiadlowski}, P., \& {Izzard}, R.~G. 2008,
  \aap, 484, 831, \dodoi{10.1051/0004-6361:200809371}

\bibitem[{{Dorn-Wallenstein} \& {Levesque}(2020)}]{DW20}
{Dorn-Wallenstein}, T.~Z., \& {Levesque}, E.~M. 2020, \apj, 896, 164,
  \dodoi{10.3847/1538-4357/ab8d28}

\bibitem[{{Duquennoy} \& {Mayor}(1991)}]{DM91}
{Duquennoy}, A., \& {Mayor}, M. 1991, \aap, 500, 337

\bibitem[{{Fuller} \& {Lu}(2022)}]{FW22}
{Fuller}, J., \& {Lu}, W. 2022, arXiv e-prints, arXiv:2201.08407.
\newblock \doarXiv{2201.08407}

\bibitem[{{Gat} {et~al.}(2013){Gat}, {van Eerten}, \& {MacFadyen}}]{Gat13}
{Gat}, I., {van Eerten}, H., \& {MacFadyen}, A. 2013, \apj, 773, 2,
  \dodoi{10.1088/0004-637X/773/1/2}

\bibitem[{Gladman {et~al.}(1996)Gladman, Quinn, Nicholson, \& Rand}]{GLAD96}
Gladman, B., Quinn, D., Nicholson, P., \& Rand, R. 1996, Icarus, 122, 166,
  \dodoi{https://doi.org/10.1006/icar.1996.0117}

\bibitem[{{Gompertz} {et~al.}(2018{\natexlab{a}}){Gompertz}, {Fruchter}, \&
  {Pe'er}}]{Gomp18}
{Gompertz}, B.~P., {Fruchter}, A.~S., \& {Pe'er}, A. 2018{\natexlab{a}}, \apj,
  866, 162, \dodoi{10.3847/1538-4357/aadba8}

\bibitem[{{Gompertz} {et~al.}(2018{\natexlab{b}}){Gompertz}, {Fruchter}, \&
  {Pe'er}}]{GFP18}
---. 2018{\natexlab{b}}, \apj, 866, 162, \dodoi{10.3847/1538-4357/aadba8}

\bibitem[{{Granot} \& {Sari}(2002)}]{GS02}
{Granot}, J., \& {Sari}, R. 2002, \apj, 568, 820, \dodoi{10.1086/338966}

\bibitem[{{Han} {et~al.}(2020){Han}, {Ge}, {Chen}, \& {Chen}}]{Han20}
{Han}, Z.-W., {Ge}, H.-W., {Chen}, X.-F., \& {Chen}, H.-L. 2020, Research in
  Astronomy and Astrophysics, 20, 161, \dodoi{10.1088/1674-4527/20/10/161}

\bibitem[{{Hancock} {et~al.}(2013){Hancock}, {Gaensler}, \& {Murphy}}]{HGM13}
{Hancock}, P.~J., {Gaensler}, B.~M., \& {Murphy}, T. 2013, \apj, 776, 106,
  \dodoi{10.1088/0004-637X/776/2/106}

\bibitem[{{Heger} {et~al.}(2005){Heger}, {Woosley}, \& {Spruit}}]{Heg05}
{Heger}, A., {Woosley}, S.~E., \& {Spruit}, H.~C. 2005, \apj, 626, 350,
  \dodoi{10.1086/429868}

\bibitem[{{Hirschi} {et~al.}(2005){Hirschi}, {Meynet}, \& {Maeder}}]{HMM05}
{Hirschi}, R., {Meynet}, G., \& {Maeder}, A. 2005, \aap, 443, 581,
  \dodoi{10.1051/0004-6361:20053329}

\bibitem[{Hjorth \& Bloom(2012)}]{HB12}
Hjorth, J., \& Bloom, J.~S. 2012, Gamma-ray bursts

\bibitem[{{Hjorth} {et~al.}(2003){Hjorth}, {Sollerman}, {M{\o}ller}, {Fynbo},
  {Woosley}, {Kouveliotou}, {Tanvir}, {Greiner}, {Andersen}, {Castro-Tirado},
  {Castro Cer{\'o}n}, {Fruchter}, {Gorosabel}, {Jakobsson}, {Kaper}, {Klose},
  {Masetti}, {Pedersen}, {Pedersen}, {Pian}, {Palazzi}, {Rhoads}, {Rol}, {van
  den Heuvel}, {Vreeswijk}, {Watson}, \& {Wijers}}]{Hjorth03}
{Hjorth}, J., {Sollerman}, J., {M{\o}ller}, P., {et~al.} 2003, \nat, 423, 847,
  \dodoi{10.1038/nature01750}

\bibitem[{{Huang} {et~al.}(2021){Huang}, {Wang}, {Liu}, {Wang}, \&
  {Liang}}]{Huang21}
{Huang}, X.-L., {Wang}, Z.-R., {Liu}, R.-Y., {Wang}, X.-Y., \& {Liang}, E.-W.
  2021, \apj, 908, 225, \dodoi{10.3847/1538-4357/abd6bc}

\bibitem[{{Hwang} {et~al.}(2021){Hwang}, {Ting}, {Schlaufman}, {Zakamska}, \&
  {Wyse}}]{Hwang21}
{Hwang}, H.-C., {Ting}, Y.-S., {Schlaufman}, K.~C., {Zakamska}, N.~L., \&
  {Wyse}, R. F.~G. 2021, \mnras, 501, 4329, \dodoi{10.1093/mnras/staa3854}

\bibitem[{{Ivanova}(2002)}]{Iv02}
{Ivanova}, N. 2002, PhD thesis, University of Oxford

\bibitem[{{Izzard} {et~al.}(2004){Izzard}, {Ramirez-Ruiz}, \& {Tout}}]{Izz04}
{Izzard}, R.~G., {Ramirez-Ruiz}, E., \& {Tout}, C.~A. 2004, \mnras, 348, 1215,
  \dodoi{10.1111/j.1365-2966.2004.07436.x}

\bibitem[{{Janiuk} {et~al.}(2013){Janiuk}, {Charzy{\'n}ski}, \&
  {Bejger}}]{Jan13}
{Janiuk}, A., {Charzy{\'n}ski}, S., \& {Bejger}, M. 2013, \aap, 560, A25,
  \dodoi{10.1051/0004-6361/201322165}

\bibitem[{{Janiuk} \& {Proga}(2008)}]{JP08}
{Janiuk}, A., \& {Proga}, D. 2008, \apj, 675, 519, \dodoi{10.1086/526511}

\bibitem[{{Kangas} \& {Fruchter}(2021)}]{KF21}
{Kangas}, T., \& {Fruchter}, A.~S. 2021, \apj, 911, 14,
  \dodoi{10.3847/1538-4357/abe76b}

\bibitem[{{King} \& {Pringle}(2021)}]{KP21}
{King}, A.~R., \& {Pringle}, J.~E. 2021, arXiv e-prints, arXiv:2107.12384.
\newblock \doarXiv{2107.12384}

\bibitem[{{Kistler} {et~al.}(2009){Kistler}, {Y{\"u}ksel}, {Beacom}, {Hopkins},
  \& {Wyithe}}]{Kist09}
{Kistler}, M.~D., {Y{\"u}ksel}, H., {Beacom}, J.~F., {Hopkins}, A.~M., \&
  {Wyithe}, J. S.~B. 2009, \apjl, 705, L104,
  \dodoi{10.1088/0004-637X/705/2/L104}

\bibitem[{{Kistler} {et~al.}(2008){Kistler}, {Y{\"u}ksel}, {Beacom}, \&
  {Stanek}}]{Kist08}
{Kistler}, M.~D., {Y{\"u}ksel}, H., {Beacom}, J.~F., \& {Stanek}, K.~Z. 2008,
  \apjl, 673, L119, \dodoi{10.1086/527671}

\bibitem[{{Klencki} {et~al.}(2021){Klencki}, {Nelemans}, {Istrate}, \&
  {Chruslinska}}]{Klen21}
{Klencki}, J., {Nelemans}, G., {Istrate}, A.~G., \& {Chruslinska}, M. 2021,
  \aap, 645, A54, \dodoi{10.1051/0004-6361/202038707}

\bibitem[{{Kobayashi} {et~al.}(1997){Kobayashi}, {Piran}, \& {Sari}}]{KPS97}
{Kobayashi}, S., {Piran}, T., \& {Sari}, R. 1997, \apj, 490, 92,
  \dodoi{10.1086/512791}

\bibitem[{{Komissarov} \& {Barkov}(2007)}]{KB07}
{Komissarov}, S.~S., \& {Barkov}, M.~V. 2007, \mnras, 382, 1029,
  \dodoi{10.1111/j.1365-2966.2007.12485.x}

\bibitem[{{Komissarov} \& {Barkov}(2009)}]{KB09}
---. 2009, \mnras, 397, 1153, \dodoi{10.1111/j.1365-2966.2009.14831.x}

\bibitem[{{Kouveliotou} {et~al.}(1993){Kouveliotou}, {Meegan}, {Fishman},
  {Bhat}, {Briggs}, {Koshut}, {Paciesas}, \& {Pendleton}}]{Kouv93}
{Kouveliotou}, C., {Meegan}, C.~A., {Fishman}, G.~J., {et~al.} 1993, \apjl,
  413, L101, \dodoi{10.1086/186969}

\bibitem[{{Kumar} {et~al.}(2008{\natexlab{a}}){Kumar}, {Narayan}, \&
  {Johnson}}]{KNJ08a}
{Kumar}, P., {Narayan}, R., \& {Johnson}, J.~L. 2008{\natexlab{a}}, Science,
  321, 376, \dodoi{10.1126/science.1159003}

\bibitem[{{Kumar} {et~al.}(2008{\natexlab{b}}){Kumar}, {Narayan}, \&
  {Johnson}}]{KNJ08b}
---. 2008{\natexlab{b}}, \mnras, 388, 1729,
  \dodoi{10.1111/j.1365-2966.2008.13493.x}

\bibitem[{{Laplace} {et~al.}(2020){Laplace}, {G{\"o}tberg}, {de Mink},
  {Justham}, \& {Farmer}}]{Lap20}
{Laplace}, E., {G{\"o}tberg}, Y., {de Mink}, S.~E., {Justham}, S., \& {Farmer},
  R. 2020, \aap, 637, A6, \dodoi{10.1051/0004-6361/201937300}

\bibitem[{{Levine} {et~al.}(2021){Levine}, {Dainotti}, {Zvonarek}, {Fraija},
  {Warren}, {Chandra}, \& {Lloyd-Ronning}}]{Lev21}
{Levine}, D., {Dainotti}, M., {Zvonarek}, K.~J., {et~al.} 2021, arXiv e-prints,
  arXiv:2111.10428.
\newblock \doarXiv{2111.10428}

\bibitem[{Li(2007)}]{Li07}
Li, L.-X. 2007, Monthly Notices of the Royal Astronomical Society: Letters,
  374, L20, \dodoi{10.1111/j.1745-3933.2006.00256.x}

\bibitem[{{Lines} {et~al.}(2015){Lines}, {Leinhardt}, {Baruteau},
  {Paardekooper}, \& {Carter}}]{Lines15}
{Lines}, S., {Leinhardt}, Z.~M., {Baruteau}, C., {Paardekooper}, S.~J., \&
  {Carter}, P.~J. 2015, \aap, 582, A5, \dodoi{10.1051/0004-6361/201526295}

\bibitem[{{Liu} {et~al.}(2021){Liu}, {Meynet}, \& {Bromm}}]{Liu21}
{Liu}, B., {Meynet}, G., \& {Bromm}, V. 2021, \mnras, 501, 643,
  \dodoi{10.1093/mnras/staa3671}

\bibitem[{{Lloyd} {et~al.}(2000){Lloyd}, {Petrosian}, \& {Mallozzi}}]{LPM00}
{Lloyd}, N.~M., {Petrosian}, V., \& {Mallozzi}, R.~S. 2000, \apj, 534, 227,
  \dodoi{10.1086/308742}

\bibitem[{{Lloyd-Ronning} {et~al.}(2020){Lloyd-Ronning}, {Hurtado}, {Aykutalp},
  {Johnson}, \& {Ceccobello}}]{LR20}
{Lloyd-Ronning}, N., {Hurtado}, V.~U., {Aykutalp}, A., {Johnson}, J., \&
  {Ceccobello}, C. 2020, \mnras, 494, 4371, \dodoi{10.1093/mnras/staa1057}

\bibitem[{{Lloyd-Ronning} {et~al.}(2018){Lloyd-Ronning}, {Lei}, \&
  {Xie}}]{LR18}
{Lloyd-Ronning}, N., {Lei}, W.-h., \& {Xie}, W. 2018, \mnras, 478, 3525,
  \dodoi{10.1093/mnras/sty1030}

\bibitem[{{Lloyd-Ronning} {et~al.}(2016){Lloyd-Ronning}, {Dolence}, \&
  {Fryer}}]{LR16}
{Lloyd-Ronning}, N.~M., {Dolence}, J.~C., \& {Fryer}, C.~L. 2016, \mnras, 461,
  1045, \dodoi{10.1093/mnras/stw1366}

\bibitem[{{Lloyd-Ronning} {et~al.}(2019{\natexlab{a}}){Lloyd-Ronning}, {Fryer},
  {Miller}, {Prasad}, {Torres}, \& {Martin}}]{LR19bz}
{Lloyd-Ronning}, N.~M., {Fryer}, C., {Miller}, J.~M., {et~al.}
  2019{\natexlab{a}}, \mnras, 485, 203, \dodoi{10.1093/mnras/stz390}

\bibitem[{{Lloyd-Ronning} \& {Fryer}(2017)}]{LR17}
{Lloyd-Ronning}, N.~M., \& {Fryer}, C.~L. 2017, \mnras, 467, 3413,
  \dodoi{10.1093/mnras/stx313}

\bibitem[{{Lloyd-Ronning} {et~al.}(2019{\natexlab{b}}){Lloyd-Ronning},
  {Gompertz}, {Pe'er}, {Dainotti}, \& {Fruchter}}]{LR19}
{Lloyd-Ronning}, N.~M., {Gompertz}, B., {Pe'er}, A., {Dainotti}, M., \&
  {Fruchter}, A. 2019{\natexlab{b}}, \apj, 871, 118,
  \dodoi{10.3847/1538-4357/aaf6ac}

\bibitem[{{Lyman} {et~al.}(2017){Lyman}, {Levan}, {Tanvir}, {Fynbo}, {McGuire},
  {Perley}, {Angus}, {Bloom}, {Conselice}, {Fruchter}, {Hjorth}, {Jakobsson},
  \& {Starling}}]{Ly17}
{Lyman}, J.~D., {Levan}, A.~J., {Tanvir}, N.~R., {et~al.} 2017, \mnras, 467,
  1795, \dodoi{10.1093/mnras/stx220}

\bibitem[{{MacFadyen} \& {Woosley}(1999)}]{MW99}
{MacFadyen}, A.~I., \& {Woosley}, S.~E. 1999, \apj, 524, 262,
  \dodoi{10.1086/307790}

\bibitem[{{Mallozzi} {et~al.}(1995){Mallozzi}, {Paciesas}, {Pendleton},
  {Briggs}, {Preece}, {Meegan}, \& {Fishman}}]{Mal95}
{Mallozzi}, R.~S., {Paciesas}, W.~S., {Pendleton}, G.~N., {et~al.} 1995, \apj,
  454, 597, \dodoi{10.1086/176513}

\bibitem[{{Marongiu} {et~al.}(2021){Marongiu}, {Guidorzi}, {Stratta}, {Gomboc},
  {Jordana-Mitjans}, {Dichiara}, {Kobayashi}, {Kopac}, \& {Mundell}}]{Mar21}
{Marongiu}, M., {Guidorzi}, C., {Stratta}, G., {et~al.} 2021, arXiv e-prints,
  arXiv:2111.00359.
\newblock \doarXiv{2111.00359}

\bibitem[{{Mathieu} \& {Geller}(2009)}]{MG09}
{Mathieu}, R.~D., \& {Geller}, A.~M. 2009, \nat, 462, 1032,
  \dodoi{10.1038/nature08568}

\bibitem[{{Mazzola} {et~al.}(2020){Mazzola}, {Badenes}, {Moe}, {Koposov},
  {Kounkel}, {Kratter}, {Covey}, {Walker}, {Thompson}, {Andrews}, {Freeman},
  {Anguiano}, {Carlberg}, {De Lee}, {Frinchaboy}, {Lewis}, {Majewski},
  {Nidever}, {Nitschelm}, {Price-Whelan}, {Roman-Lopes}, {Stassun}, \&
  {Troup}}]{Mazz20}
{Mazzola}, C.~N., {Badenes}, C., {Moe}, M., {et~al.} 2020, \mnras, 499, 1607,
  \dodoi{10.1093/mnras/staa2859}

\bibitem[{{McKinney}(2005)}]{mck05}
{McKinney}, J.~C. 2005, \apjl, 630, L5, \dodoi{10.1086/468184}

\bibitem[{{Nakar} \& {Granot}(2007)}]{NG07}
{Nakar}, E., \& {Granot}, J. 2007, \mnras, 380, 1744,
  \dodoi{10.1111/j.1365-2966.2007.12245.x}

\bibitem[{{Nakar} \& {Piran}(2002)}]{NP02}
{Nakar}, E., \& {Piran}, T. 2002, \apjl, 572, L139, \dodoi{10.1086/341748}

\bibitem[{{Olofsson} {et~al.}(2015){Olofsson}, {Vlemmings}, {Maercker},
  {Humphreys}, {Lindqvist}, {Nyman}, \& {Ramstedt}}]{Olof15}
{Olofsson}, H., {Vlemmings}, W., {Maercker}, M., {et~al.} 2015, Astronomical
  Society of the Pacific Conference Series, Vol. 499, {An ALMA View of the
  Complex Circumstellar Environment of the Post-AGB Object HD 101584}, ed.
  D.~{Iono}, K.~{Tatematsu}, A.~{Wootten}, \& L.~{Testi}, 319

\bibitem[{{Pacholczyk}(1970)}]{Pach70}
{Pacholczyk}, A.~G. 1970, {Radio astrophysics. Nonthermal processes in galactic
  and extragalactic sources}

\bibitem[{{Perley} {et~al.}(2016){Perley}, {Niino}, {Tanvir}, {Vergani}, \&
  {Fynbo}}]{Per16}
{Perley}, D.~A., {Niino}, Y., {Tanvir}, N.~R., {Vergani}, S.~D., \& {Fynbo},
  J.~P.~U. 2016, \ssr, 202, 111, \dodoi{10.1007/s11214-016-0237-4}

\bibitem[{{Piran}(2004)}]{Pir04}
{Piran}, T. 2004, Reviews of Modern Physics, 76, 1143,
  \dodoi{10.1103/RevModPhys.76.1143}

\bibitem[{{Podsiadlowski} {et~al.}(2010){Podsiadlowski}, {Ivanova}, {Justham},
  \& {Rappaport}}]{Pod10}
{Podsiadlowski}, P., {Ivanova}, N., {Justham}, S., \& {Rappaport}, S. 2010,
  \mnras, 406, 840, \dodoi{10.1111/j.1365-2966.2010.16751.x}

\bibitem[{{Podsiadlowski} {et~al.}(2004){Podsiadlowski}, {Mazzali}, {Nomoto},
  {Lazzati}, \& {Cappellaro}}]{Pod04}
{Podsiadlowski}, P., {Mazzali}, P.~A., {Nomoto}, K., {Lazzati}, D., \&
  {Cappellaro}, E. 2004, \apjl, 607, L17, \dodoi{10.1086/421347}

\bibitem[{{Price} \& {Rosswog}(2006)}]{Pr06}
{Price}, D.~J., \& {Rosswog}, S. 2006, Science, 312, 719,
  \dodoi{10.1126/science.1125201}

\bibitem[{{Proga} \& {Begelman}(2003)}]{PB03}
{Proga}, D., \& {Begelman}, M.~C. 2003, \apj, 592, 767, \dodoi{10.1086/375773}

\bibitem[{{Rosas-Guevara} {et~al.}(2015){Rosas-Guevara}, {Bower}, {Schaye},
  {Furlong}, {Frenk}, {Booth}, {Crain}, {Dalla Vecchia}, {Schaller}, \&
  {Theuns}}]{RG15}
{Rosas-Guevara}, Y.~M., {Bower}, R.~G., {Schaye}, J., {et~al.} 2015, \mnras,
  454, 1038, \dodoi{10.1093/mnras/stv2056}

\bibitem[{{Rosen} {et~al.}(2012){Rosen}, {Krumholz}, \& {Ramirez-Ruiz}}]{Ros12}
{Rosen}, A.~L., {Krumholz}, M.~R., \& {Ramirez-Ruiz}, E. 2012, \apj, 748, 97,
  \dodoi{10.1088/0004-637X/748/2/97}

\bibitem[{{Sana} {et~al.}(2012){Sana}, {de Mink}, {de Koter}, {Langer},
  {Evans}, {Gieles}, {Gosset}, {Izzard}, {Le Bouquin}, \& {Schneider}}]{Sana12}
{Sana}, H., {de Mink}, S.~E., {de Koter}, A., {et~al.} 2012, Science, 337, 444,
  \dodoi{10.1126/science.1223344}

\bibitem[{{Sana} {et~al.}(2013){Sana}, {de Koter}, {de Mink}, {Dunstall},
  {Evans}, {H{\'e}nault-Brunet}, {Ma{\'\i}z Apell{\'a}niz},
  {Ram{\'\i}rez-Agudelo}, {Taylor}, {Walborn}, {Clark}, {Crowther}, {Herrero},
  {Gieles}, {Langer}, {Lennon}, \& {Vink}}]{Sana13}
{Sana}, H., {de Koter}, A., {de Mink}, S.~E., {et~al.} 2013, \aap, 550, A107,
  \dodoi{10.1051/0004-6361/201219621}

\bibitem[{{Sari} {et~al.}(1998){Sari}, {Piran}, \& {Narayan}}]{SPN98}
{Sari}, R., {Piran}, T., \& {Narayan}, R. 1998, \apjl, 497, L17,
  \dodoi{10.1086/311269}

\bibitem[{{Schr{\o}der} {et~al.}(2021){Schr{\o}der}, {MacLeod}, {Ramirez-Ruiz},
  {Mandel}, {Fragos}, {Loeb}, \& {Everson}}]{Schr21}
{Schr{\o}der}, S.~L., {MacLeod}, M., {Ramirez-Ruiz}, E., {et~al.} 2021, arXiv
  e-prints, arXiv:2107.09675.
\newblock \doarXiv{2107.09675}

\bibitem[{{Sundqvist} {et~al.}(2013){Sundqvist}, {Sim{\'o}n-D{\'\i}az}, {Puls},
  \& {Markova}}]{Sund13}
{Sundqvist}, J.~O., {Sim{\'o}n-D{\'\i}az}, S., {Puls}, J., \& {Markova}, N.
  2013, \aap, 559, L10, \dodoi{10.1051/0004-6361/201322761}

\bibitem[{{Tandon} \& {Lloyd-Ronning}(2021)}]{TLR21}
{Tandon}, C., \& {Lloyd-Ronning}, N. 2021, Research Notes of the American
  Astronomical Society, 5, 184, \dodoi{10.3847/2515-5172/ac1a7c}

\bibitem[{{Tchekhovskoy} \& {Giannios}(2015)}]{TG15}
{Tchekhovskoy}, A., \& {Giannios}, D. 2015, \mnras, 447, 327,
  \dodoi{10.1093/mnras/stu2229}

\bibitem[{{Tchekhovskoy} {et~al.}(2010){Tchekhovskoy}, {Narayan}, \&
  {McKinney}}]{TNM10}
{Tchekhovskoy}, A., {Narayan}, R., \& {McKinney}, J.~C. 2010, \apj, 711, 50,
  \dodoi{10.1088/0004-637X/711/1/50}

\bibitem[{{van den Heuvel} \& {Yoon}(2007)}]{vdHy07}
{van den Heuvel}, E.~P.~J., \& {Yoon}, S.~C. 2007, \apss, 311, 177,
  \dodoi{10.1007/s10509-007-9583-8}

\bibitem[{{van Eerten} \& {MacFadyen}(2013)}]{vE13}
{van Eerten}, H., \& {MacFadyen}, A. 2013, \apj, 767, 141,
  \dodoi{10.1088/0004-637X/767/2/141}

\bibitem[{Vianello {et~al.}(2018)Vianello, Gill, Granot, Omodei, Cohen-Tanugi,
  \& Longo}]{Via18}
Vianello, G., Gill, R., Granot, J., {et~al.} 2018, The Astrophysical Journal,
  864, 163, \dodoi{10.3847/1538-4357/aad6ea}

\bibitem[{{Wang} {et~al.}(2020){Wang}, {Zou}, {Liu}, {Liao}, {Liu}, {Chai}, \&
  {Xia}}]{Wang20}
{Wang}, F., {Zou}, Y.-C., {Liu}, F., {et~al.} 2020, \apj, 893, 77,
  \dodoi{10.3847/1538-4357/ab0a86}

\bibitem[{Woosley \& Bloom(2006)}]{WB06}
Woosley, S., \& Bloom, J. 2006, Annu. Rev. Astron. Astrophys., 44, 507

\bibitem[{{Woosley}(1993)}]{Woos93}
{Woosley}, S.~E. 1993, \apj, 405, 273, \dodoi{10.1086/172359}

\bibitem[{{Woosley} \& {Heger}(2006{\natexlab{a}})}]{WH06}
{Woosley}, S.~E., \& {Heger}, A. 2006{\natexlab{a}}, \apj, 637, 914,
  \dodoi{10.1086/498500}

\bibitem[{{Woosley} \& {Heger}(2006{\natexlab{b}})}]{Heg06}
---. 2006{\natexlab{b}}, \apj, 637, 914, \dodoi{10.1086/498500}

\bibitem[{{Yoon} \& {Langer}(2005)}]{YL05}
{Yoon}, S.-C., \& {Langer}, N. 2005, \aap, 443, 643,
  \dodoi{10.1051/0004-6361:20054030}

\bibitem[{{Yoon} {et~al.}(2006){Yoon}, {Langer}, \& {Norman}}]{Yoon06}
{Yoon}, S.-C., {Langer}, N., \& {Norman}, C. 2006, \aap, 460, 199,
  \dodoi{10.1051/0004-6361:20065912}

\bibitem[{{Y{\"u}ksel} {et~al.}(2008){Y{\"u}ksel}, {Kistler}, {Beacom}, \&
  {Hopkins}}]{Yuk08}
{Y{\"u}ksel}, H., {Kistler}, M.~D., {Beacom}, J.~F., \& {Hopkins}, A.~M. 2008,
  \apjl, 683, L5, \dodoi{10.1086/591449}

\bibitem[{{Zapartas} {et~al.}(2021){Zapartas}, {de Mink}, {Justham}, {Smith},
  {Renzo}, \& {de Koter}}]{Zap21}
{Zapartas}, E., {de Mink}, S.~E., {Justham}, S., {et~al.} 2021, \aap, 645, A6,
  \dodoi{10.1051/0004-6361/202037744}

\bibitem[{{Zhang} {et~al.}(2021){Zhang}, {Zhang}, {Huang}, {Song}, {Zheng},
  {Li}, {Li}, \& {Su}}]{Zhang21}
{Zhang}, K., {Zhang}, Z.~B., {Huang}, Y.~F., {et~al.} 2021, \mnras, 503, 3262,
  \dodoi{10.1093/mnras/stab465}

\bibitem[{{Zou} {et~al.}(2019){Zou}, {Zhou}, {Xie}, {Zhang}, {L{\"u}}, {Zhong},
  {Wang}, \& {Liang}}]{Zou19}
{Zou}, L., {Zhou}, Z.-M., {Xie}, L., {et~al.} 2019, \apj, 877, 153,
  \dodoi{10.3847/1538-4357/ab17dc}

\end{thebibliography}
\bibliographystyle{aasjournal}



\end{document}